%% file: main.tex
\documentclass[sigconf, nonacm]{acmart}

\usepackage[a-2b]{pdfx}
\usepackage[T1]{fontenc}

\newcommand\vldbdoi{XX.XX/XXX.XX}
\newcommand\vldbpages{XXX-XXX}
\newcommand\vldbvolume{19}
\newcommand\vldbissue{11}
\newcommand\vldbyear{2026}
\newcommand\vldbauthors{\authors}
\newcommand\vldbtitle{\shorttitle}
\newcommand\vldbavailabilityurl{https://github.com/PKU-DAIR/MFTune}
\newcommand\vldbpagestyle{empty}

\usepackage{enumitem}
\usepackage[normalem]{ulem}
\usepackage{subcaption}
\usepackage{xspace}
\usepackage{bm}
\usepackage{tabularx}
\usepackage{algorithm}
\usepackage{algorithmic}
\usepackage{tikz}
\usepackage{textcomp}
\usepackage[most]{tcolorbox}
\usepackage{multibib}
\usepackage{makecell}
\usepackage{multirow}
\usepackage{cuted}
\usepackage{array}
\newcolumntype{C}[1]{>{\centering\arraybackslash}m{#1}}

\newcites{resp}{References for Response Letter}

\tcbset{
  width=\columnwidth,
  boxrule=0pt,
  colback=lightgray!30,
  arc=0pt,
  auto outer arc,
  left=2pt,
  right=2pt,
  boxsep=3pt
}

 \newcommand{\revise}[1]{#1}

\newcommand{\sys}{MFTune\xspace}

\begin{document}

\title{\sys: An Efficient Multi-fidelity Framework for Spark SQL Configuration Tuning}


\author{Beicheng Xu}
\authornote{Both authors contributed equally to this research.}
\authornote{All authors are with the School of Computer Science \& Key Lab of High Confidence Software Technologies (MOE) \& Beijing Key Laboratory of Software and Hardware Cooperative Artificial Intelligence Systems, Peking University, Beijing, China.}
\affiliation{%
  \institution{Peking University}
  \city{Beijing}
  \country{China}}
\email{beichengxu@stu.pku.edu.cn}
\author{Lingching Tung}
\authornotemark[1]
\authornotemark[2]
\affiliation{%
  \institution{Peking University}
  \city{Beijing}
  \country{China}}
\email{lingchingtung@stu.pku.edu.cn}
\author{Yuchen Wang}
\authornotemark[2]
\affiliation{%
  \institution{Peking University}
  \city{Beijing}
  \country{China}}
\email{wychen@stu.pku.edu.cn}
\author{Yupeng Lu}
\authornotemark[2]
\affiliation{%
  \institution{Peking University}
  \city{Beijing}
  \country{China}}
\email{xinkelyp@pku.edu.cn}
\author{Bin Cui}
\authornotemark[2]
\affiliation{%
  \institution{Peking University}
  \city{Beijing}
  \country{China}}
\email{bin.cui@pku.edu.cn}

\begin{abstract}
Apache Spark SQL is a cornerstone of modern big data analytics.
However, optimizing Spark SQL performance is challenging due to its vast configuration space and the prohibitive cost of evaluating massive workloads.
Existing tuning methods predominantly rely on full-fidelity evaluations, which are extremely time-consuming, often leading to suboptimal performance within practical budgets.
While multi-fidelity optimization offers a potential solution, directly applying standard techniques—such as data volume reduction or early stopping—proves ineffective for Spark SQL as they fail to preserve performance correlations or represent true system bottlenecks.
To address these challenges, we propose \sys, an efficient multi-fidelity framework that introduces a query-based fidelity partitioning strategy, utilizing representative SQL subsets to provide accurate, low-cost proxies. To navigate the huge search space, \sys incorporates a density-based optimization mechanism for automated knob and range compression, alongside an adapted transfer learning approach and a two-phase warm start to further accelerate the tuning process.
Experimental results on TPC-H and TPC-DS benchmarks demonstrate that MFTune significantly outperforms five state-of-the-art tuning methods, identifying superior configurations within practical time constraints.
\end{abstract}

\maketitle

\pagestyle{\vldbpagestyle}
\begingroup\small\noindent\raggedright\textbf{PVLDB Reference Format:}\\
\vldbauthors. \vldbtitle. PVLDB, \vldbvolume(\vldbissue): \vldbpages, \vldbyear.\\
\href{https://doi.org/\vldbdoi}{doi:\vldbdoi}
\endgroup
\begingroup
\renewcommand\thefootnote{}\footnote{\noindent
This work is licensed under the Creative Commons BY-NC-ND 4.0 International License. Visit \url{https://creativecommons.org/licenses/by-nc-nd/4.0/} to view a copy of this license. For any use beyond those covered by this license, obtain permission by emailing \href{mailto:info@vldb.org}{info@vldb.org}. Copyright is held by the owner/author(s). Publication rights licensed to the VLDB Endowment. \\
\raggedright Proceedings of the VLDB Endowment, Vol. \vldbvolume, No. \vldbissue\ %
ISSN 2150-8097. \\
\href{https://doi.org/\vldbdoi}{doi:\vldbdoi} \\
}\addtocounter{footnote}{-1}\endgroup

\ifdefempty{\vldbavailabilityurl}{}{
\vspace{.3cm}
\begingroup\small\noindent\raggedright\textbf{PVLDB Artifact Availability:}\\
The source code, data, and/or other artifacts have been made available at \url{\vldbavailabilityurl}.
\endgroup
}

\input{tex/introduction}
\input{tex/related_work}
\input{tex/preliminary}

\input{tex/overview}

\input{tex/method}
\input{tex/experiment}

\section{Conclusion}
In this paper, we propose \sys, a multi-fidelity framework that overcomes the prohibitive costs and high dimensionality of Spark SQL tuning through query-based fidelity partitioning and density-based search space compression. By integrating a two-phase warm start and adapted transfer learning, MFTune achieves a superior balance between historical knowledge reuse and efficient exploration. Experimental results on TPC-H and TPC-DS benchmarks confirm that MFTune consistently delivers optimal performance across all evaluated settings, significantly outperforming five state-of-the-art methods with negligible computational overhead.


\begin{acks}
  This work is supported by the National Natural Science Foundation of
  China (U23B2048, U22B2037) and the ``Fundamental Research Funds for the Central Universities, Peking University''. Bin Cui is the corresponding author.
\end{acks}


\bibliographystyle{ACM-Reference-Format}
\bibliography{sample}

\end{document}

%% file: tex/introduction.tex
\section{Introduction}
\label{sec:introduction}

\begin{figure}[t]
    \centering
    \begin{subfigure}[b]{0.50\linewidth}
        \centering
        \includegraphics[width=\linewidth]{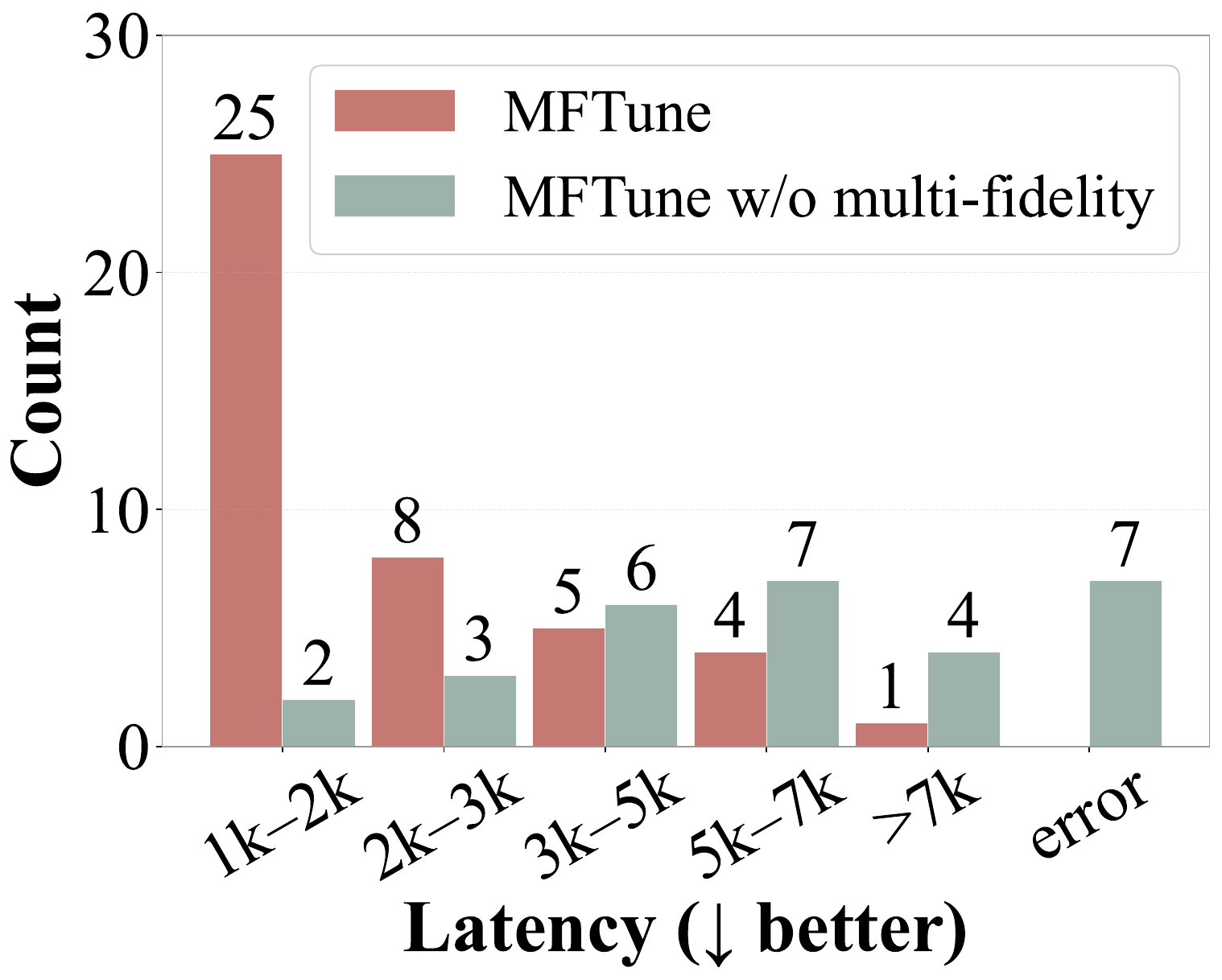}
        \caption{Performance distribution.}
        \label{fig:motivation_of_mf}
    \end{subfigure}
    \hfill
    \begin{subfigure}[b]{0.49\linewidth}
        \centering
        \includegraphics[width=\linewidth]{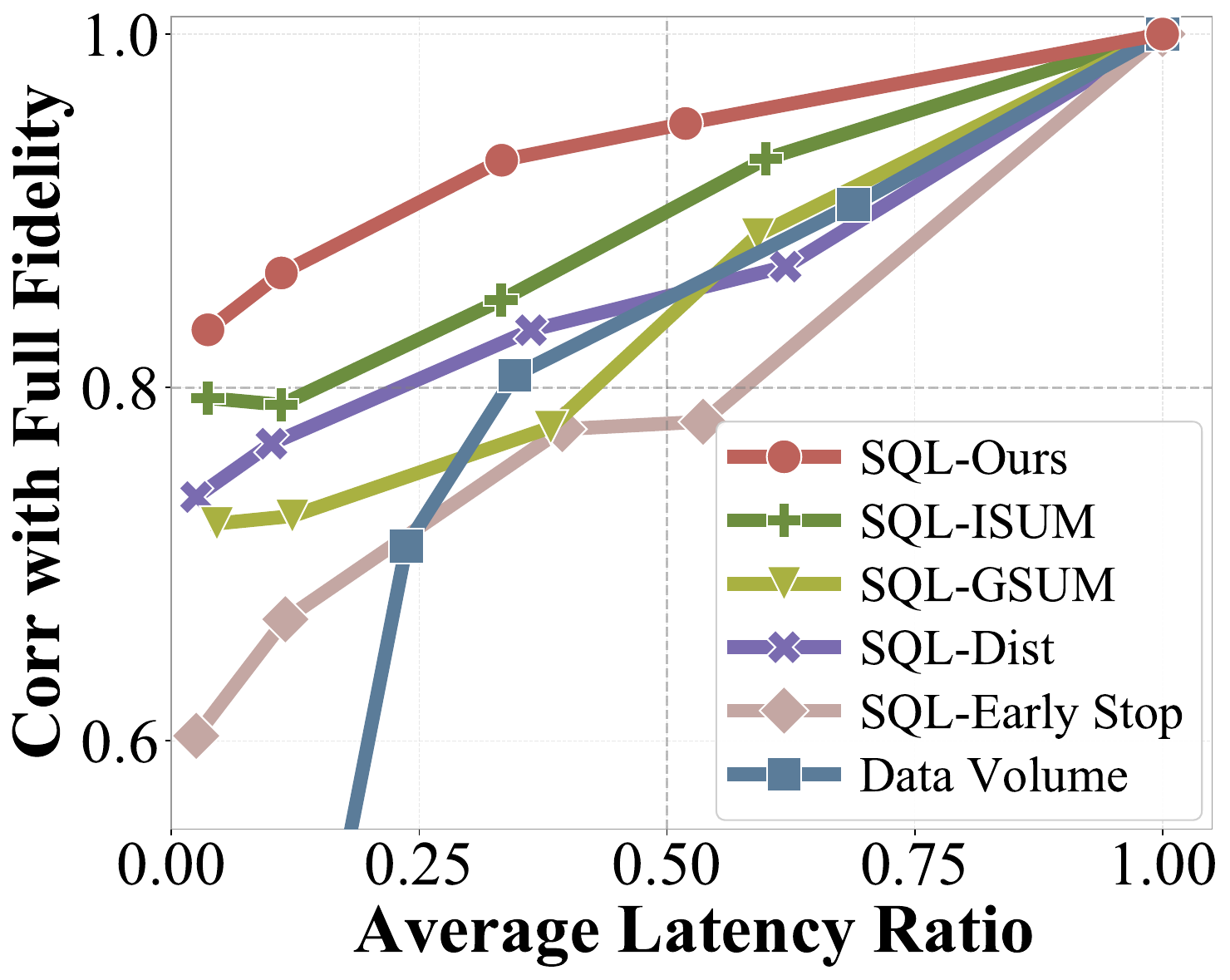}
        \caption{Fidelity correlation.}
        \label{fig:partition}
    \end{subfigure}

    \caption[Analysis of multi-fidelity mechanism on TPC-DS (600GB)]{\textbf{Analysis of multi-fidelity mechanism on TPC-DS (600GB).}
    (a) Latency distribution of configurations evaluated within 48 hours.
    (b) Fidelity correlation. \revise{We evaluate six proxies: 1) SQL-Ours, selecting SQL subsets from $\frac{1}{27}$ to $1$ fidelity based on Section~\ref{sec:fidelity_partition}; 2) SQL-ISUM, 3) SQL-GSUM, and 4) SQL-Dist, adapted from ISUM~\cite{siddiqui2022isum}, GSUM~\cite{deep2020comprehensive}, and the distance-based method~\cite{chaudhuri2002compressing} to our setting, respectively, detailed in Section~\ref{sec:exp_analysis_of_MFO}; 5) SQL-Early Stop, executing prefixes from $\frac{1}{27}$ to $1$ of the total SQLs; and 6) Data Volume, scaling the dataset across \{30,100,200,400,600\} GB. 
    Each point reports the Kendall correlation between proxy and full-fidelity latency versus the average latency ratio, computed over 50 configurations.}}
    \label{fig:budget_allocation}
\end{figure}

The recent explosion of data has catalyzed the widespread deployment of large-scale query systems, such as Hive~\cite{hive_thusoo2009hive}, Presto~\cite{presto_sethi2019presto}, and Spark SQL~\cite{armbrust2015spark}. 
Among these, Spark SQL has emerged as a premier choice for industry-scale data processing due to its robust integration with the Spark ecosystem~\cite{spark_zaharia2010spark} and its superior performance in handling complex analytical workloads~\cite{jin2025postman}. 
In real-world scenarios, Spark SQL workloads typically have two characteristics.
  First, they often include multiple queries, as production data processing tasks are commonly structured as multi-stage workflows, such as Extract, Load, and Transform pipelines in data warehousing~\cite{kimball2013data} and feature generation for machine
  learning~\cite{huyen2022designing}.  
  Second, they typically process large volumes of data, spanning hundreds of gigabytes to several terabytes or more~\cite{warren2015big}.
The combination of multiple queries and massive datasets makes Spark SQL workloads highly resource-intensive, often requiring hours or even days to complete.
  Meanwhile, the efficiency of Spark SQL is heavily dependent on more than 200 configuration knobs that govern system behaviors like memory management and shuffle partitioning~\cite{spark_configuration}.
  Because default settings are often suboptimal, each evaluation is costly, and the configuration space is prohibitively large, manual tuning is both time-consuming and far from optimal.

Therefore, various configuration tuning methods have been proposed to find optimal configurations for Spark SQL workloads automatically~\cite{xin2022locat,shen2023rover,li2023towards,li2025loftune,chen2024tie,fekry2020tune}.
These methods typically adopt the Bayesian Optimization (BO) framework~\cite{snoek2012practical,hutter2011sequential,frazzetto2025graph} to sample promising configurations, iteratively replay the workload, and model the relationship between system performance and parameters.
However, these methods suffer from several fundamental challenges when applied to Spark SQL, as follows:

\textbf{C1. Prohibitive evaluation costs.}
Given the long-running nature of Spark SQL workloads, evaluating configurations is highly expensive.
However, existing tuning methods often rely on full-fidelity evaluations (executing the entire original workload)~\cite{fekry2020tune,li2025loftune,li2023towards}, which limits the number of configurations that can be explored within a finite tuning budget.
A potential solution to mitigate this bottleneck is to adopt Multi-Fidelity Optimization (MFO)~\cite{li2021mfes,li2018hyperband,falkner2018bohb}.
MFO leverages low-fidelity evaluations, which are faster but approximate the full-fidelity results, to efficiently filter out poor configurations and guide the search toward optimal regions.
Our preliminary analysis in Figure~\ref{fig:motivation_of_mf} demonstrates the impact of this mechanism: within a 48-hour budget, \sys successfully evaluated 43 configurations with MFO.
In contrast, a full-fidelity variant (\sys w/o multi-fidelity) evaluated only 29 configurations in the same time frame, with the majority falling into higher latency buckets or even resulting in execution errors. 
This comparison illustrates that relying solely on full-fidelity evaluations makes it infeasible to explore the vast search space effectively within practical budgets.

\textbf{C2. Inapplicability of directly applying existing MFO methods.}
The core challenge of MFO lies in ensuring that the low-fidelity proxies preserve the relative ranking of configurations observed under full fidelity.
However, existing standard fidelity-partitioning techniques used in Automated Machine Learning (AutoML), such as training on a subset of data~\cite{li2021mfes, klein2017fast, li2022hyper} or for fewer epochs~\cite{jiang2024efficient,li2022hyper,falkner2018bohb}, do not transfer effectively to Spark SQL.
\revise{As illustrated in Figure~\ref{fig:partition}, generic low-fidelity proxies in AutoML fail to maintain a reliable correlation with full-fidelity performance in Spark SQL.}
Reducing data volume leads to a sharp decline in correlation because smaller datasets fail to trigger the true bottlenecks at full scale.
\revise{Beyond generic proxies, we also adapt DBMS workload-compression methods~\cite{siddiqui2022isum,deep2020comprehensive,chaudhuri2002compressing}. 
However, the same figure shows that they still provide weak correlation with full-fidelity as they are designed for different downstream objectives like index tuning rather than preserving configuration rankings for Spark SQL MFO.}
In contrast, our proposed SQL Selection maintains a high Kendall-tau correlation (>0.8) even at extremely low fidelity levels.
These findings demonstrate that customizing fidelity partitioning for the Spark SQL domain is essential, as neither generic AutoML proxies nor DBMS workload-compression methods can provide the reliable lightweight feedback required for effective tuning.

\textbf{C3. Huge search space and ineffective compression.}
The search space of Spark SQL is prohibitively large for exhaustive exploration within a limited budget. 
Most existing works focus solely on reducing the number of knobs via sensitivity analysis or projection~\cite{xin2022locat,fekry2020tune,wei2025toptune,li2023towards}. 
Only a few methods~\cite{zhang2023efficient,sun2025rabbit} have attempted to compress the value range. 
However, their bucket- or boundary-based strategies are sensitive to outliers and may miss promising but underexplored regions. 
Consequently, robust and fine-grained search-space compression remains an open challenge.

To address these challenges, we propose \sys, an efficient multi-fidelity framework tailored for Spark SQL configuration tuning.
Our key contributions are summarized as follows:

\begin{itemize}[leftmargin=2em, topsep=2pt, partopsep=3pt, itemsep=3pt, parsep=0pt]

\item
To the best of our knowledge, we are the first to introduce multi-fidelity optimization (MFO) into Spark SQL tuning. We further tailor transfer learning and warm-start mechanisms to accommodate varying fidelity levels.

\item We propose a query-based multi-fidelity partitioning algorithm to select representative SQL query subsets that maintain a high performance correlation with full-fidelity workloads.

\item We introduce a unified framework for knob selection and range compression based on weighted density modeling, which captures high-potential regions accurately.

\item Extensive evaluations across various benchmarks, hardwares, and data scales demonstrate the superiority of \sys over state-of-the-art baselines, yielding 25.9\%--43.1\% and 37.8\%--69.3\% latency reduction on TPC-H and TPC-DS, respectively.

\end{itemize}

%% file: tex/related_work.tex
\section{Related work}
\label{sec:related_work}


\subsection{Automatic System Configuration Tuning}
\label{sec:automatic_system_configuration_tuning}

\textbf{DBMS tuning}.
Various automatic configuration tuning methods have been proposed for Database Management Systems (DBMS)~\cite{zhao2023automatic,li2024opengauss,huang2023survey}.
Among them, Bayesian Optimization (BO)-based methods and Reinforcement Learning (RL)-based methods have been the most widely used.
BO-based methods (e.g., OtterTune~\cite{van2017automatic}, ResTune~\cite{zhang2021restune}, LlamaTune~\cite{kanellis2022llamatune}, OpAdvisor~\cite{zhang2023efficient},  TopTune~\cite{wei2025toptune}) iteratively sample configurations and replay the workload to model the relation between configuration and performance.
RL-based methods (e.g., CDBTune~\cite{zhang2019end}, QTune~\cite{li2019qtune}, SHA~\cite{li2024sha}, UDO~\cite{wang2022udo}) use RL to predict the reward of a configuration, and explore new configurations to maximize the reward, which requires more resources to converge.
\revise{
  Lero~\cite{zhu2023lero} uses a pairwise ranker to select execution plans for query optimization.
  Recent studies have also leveraged LLMs for DBMS tuning.
  DB-BERT~\cite{trummer2022db} extracts tuning hints from manuals and documents, while GPTuner~\cite{lao2025gptuner}, Rabbit~\cite{sun2025rabbit}, and AgentTune~\cite{li2025agenttune} use LLMs or LLM agents to exploit
  textual knowledge for knob selection, range pruning, and configuration optimization.
}
Another line of studies focuses on the online setting and addresses the workload shift during the tuning process~\cite{zhang2022towards, shen2025tune}.
\revise{
  In the noisy cloud environment, TUNA~\cite{freischuetz2025tuna} improves robustness by evaluating configurations across machines and adjusting noisy measurements.
}

\noindent \textbf{Spark SQL Tuning}.
While DBMS tuning methods can in principle be adapted to Spark SQL, the significantly longer query execution times in Spark SQL make methods that require hundreds of iterations~\cite{zhang2021restune,zhang2023efficient,cai2022hunter} extremely time-consuming.
Therefore, recent studies have developed Spark SQL-specific auto-tuning methods, most of which are based on BO and strive to enhance tuning efficiency to mitigate the evaluation overhead.
For example, TIE~\cite{chen2024tie} accelerates the sampling phase of BO through an early termination mechanism, thereby reducing the tuning time.
LOCAT~\cite{xin2022locat} compresses the target workload after obtaining sufficient evaluation observations of the original workload, and performs BO in a reduced search space.
To avoid tuning each workload from scratch, some methods leverage knowledge from similar historical tuning tasks~\cite{fekry2020tune,li2023towards,shen2023rover,li2025loftune}.
For example, Tuneful~\cite{fekry2020tune} employs a Multi-Task Gaussian Process to train and predict similar historical task data alongside current task observations.
Online-Tune~\cite{li2023towards} and LOFTune~\cite{li2025loftune} obtain optimal configurations from similar tasks for a warm start.
Rover~\cite{shen2023rover} accelerates tuning by weighting the BO acquisition function with historical tasks.
\revise{
  Rockhopper~\cite{zhu2025rockhopper} studies online Spark configuration tuning in production environments, focusing on robust adaptation to noisy and dynamic workloads.
}
However, all the aforementioned methods share a common limitation: they rely on complete evaluations of the workload.
Within a limited budget, it's challenging to evaluate enough configurations on complex, large-scale Spark SQL workloads, leading to insufficient search space exploration and reduced optimization efficiency.
LOCAT~\cite{xin2022locat} is the only approach that considers workload compression, but it requires extensive random search on the original workload to collect compression observations, which remains time-consuming.
Overall, efficient Spark SQL configuration tuning remains a significant challenge.

\subsection{Multi-Fidelity Optimization}
Although system parameter tuning currently relies on complete evaluations, AutoML has been exploring multi-fidelity optimization to tackle this challenge~\cite{li2018hyperband,hu2019multi,jiang2024efficient,klein2017fast,salinas2023optimizing}.
Typical MFO methods use cheap low-fidelity evaluations to guide search:
Hyperband~\cite{li2018hyperband} (HB) dynamically allocates resources to
a set of random configurations, and uses the successive halving
algorithm~\cite{jamieson2016non} to stop poor-performing configurations in advance.
BOHB~\cite{falkner2018bohb} and MFES-HB~\cite{li2021mfes} improve HB by replacing random sampling with BO.
ASHA~\cite{li2018massively}, A-BOHB~\cite{klein2020model}, and Hyper-Tune~\cite{li2022hyper} extend HB to the asynchronous case when using multiple workers.
An orthogonal line of work models the learning curves of machine learning algorithms directly~\cite{mohr2024learning}.
These studies often use neural networks to rank learning curves across different configurations~\cite{klein2017learning,wistuba2020learning}.
\revise{
  MFIX~\cite{chang2024mfix} further adapts MFO to database index advising.
  It uses a what-if optimizer interface as the low-fidelity proxy, which is not available in Spark SQL knob tuning.
}

\revise{
MFO provides an opportunity to reduce reliance on complete evaluations through cheaper proxies.
An effective low-fidelity proxy should preserve full-fidelity configuration rankings, since MFO uses it to prune poor configurations early.
However, existing MFO methods define low-fidelity proxies for their own domains and do not transfer well to Spark SQL tuning.
As shown in Figure~\ref{fig:partition}, generic proxies such as data volume reduction or early stop have low correlation with the full-fidelity performance, making them unreliable for Spark SQL configuration tuning.
}

\subsection{\texorpdfstring{\revise{Workload Compression}}{Workload Compression}}
\label{sec:workload_compression}
\revise{
The database community has long studied workload compression, which replaces a large SQL workload with a subset for downstream applications.
Chaudhuri et al.~\cite{chaudhuri2002compressing} define application-specific query distances to bound the loss of replacing one query with another for index selection and approximate query processing.
GSUM~\cite{deep2020comprehensive} constructs representative query subsets by matching feature distributions and improving feature coverage.
ISUM~\cite{siddiqui2022isum} compresses workloads for scalable index tuning by selecting queries with high utility-influence benefit, where utility estimates index-induced cost reduction and influence is derived from indexable-column features.

These studies show the potential to reduce large workloads for downstream DBMS tasks.
However, directly using them as Spark SQL fidelity proxies faces two mismatches:
First, several required signals~\cite{chaudhuri2002compressing,siddiqui2022isum}, such as hypothetical-index benefits and AQP approximation errors, are unavailable in the Spark SQL scenario.
Second, they select queries by optimizing feature coverage~\cite{deep2020comprehensive}, or query-to-query influence~\cite{chaudhuri2002compressing,siddiqui2022isum}, rather than directly optimizing the configuration-ranking consistency between the subset and the full workload.
This ranking consistency is essential in Spark SQL multi-fidelity tuning, where low-fidelity evaluations decide which configurations proceed to full-workload evaluation.
Accordingly, Figure~\ref{fig:partition} shows that even adapted workload-compression methods still provide weak correlation with full-fidelity performance.
Together with the limitations of generic MFO proxies, this underscores the need for a Spark SQL-specific fidelity partitioning strategy.

}

%% file: tex/preliminary.tex
\section{Preliminaries}

\begin{figure*}[!t]
	\centering
	\scalebox{1}[1] {
	\includegraphics[width=0.98\linewidth]{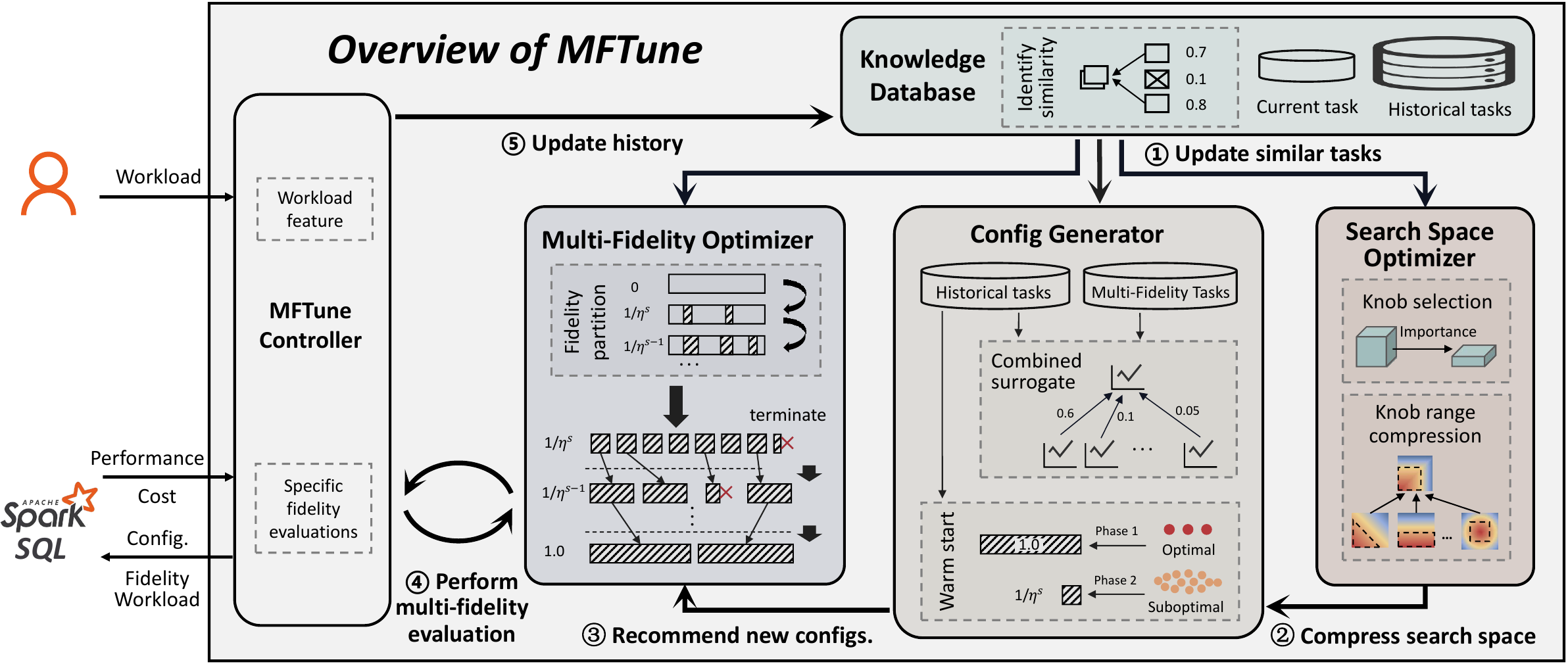}}
	\caption{Overview: architecture and workflow of \sys.}
    \label{fig:overview}
\end{figure*}

\subsection{Problem Definition}
Given a target workload $W$, which consists of multiple SQL queries over a database, the goal of Spark SQL configuration tuning is to find the optimal Spark configuration $\bm{x}^*$ that minimizes a metric:
\begin{equation}
\bm{x}^* = \arg \min_{\bm{x} \in \mathcal{X}} f(\bm{x}, W),
\end{equation}
where $\mathcal{X}$ is the Spark configuration search space and $f(\bm{x}, W)$ is the performance (in our case, latency) of $W$ under configuration $\bm{x}$.

\subsection{Bayesian Optimization Framework}

Bayesian optimization (BO)~\cite{snoek2012practical,hutter2011sequential,bergstra2011algorithms} is a popular algorithm framework that solves black-box problems with a costly objective function. 
It loops over the following steps:
1) it fits a \textit{surrogate model} $M$ based on observervations $D = \{(\bm{x}_1, y_1),...,(\bm{x}_{n-1}, y_{n-1})\}$, where $\bm{x}_i$ is the $i^{th}$ configuration and $y_i$ is the corresponding observed performance; 
2) with the surrogate $M$, BO chooses the next configuration by maximizing the \textit{acquisition function} $\bm{x}_{n}=\arg\max_{\bm{x} \in \mathcal{X}}\alpha(\bm{x}; M)$;
3) It evaluates the configuration ${\bm x}_n$ to obtain its performance $y_n$ and augments the observations by $D=D\cup\{({\bm x}_n,y_n)\}$.

\textbf{Surrogate}.
We apply the Probabilistic Random Forest~\cite{hutter2011sequential} as the surrogate.
Compared to alternatives, it has lower time complexity and shows better empirical results in real-world scenarios~\cite{zhang2022facilitating}.

\textbf{Acquisition function}.
We apply Expected Improvement (EI)~\cite{jones1998efficient} to select candidate configurations.
Given the surrogate predictive distribution $p_M(y|\bm{x})$, EI measures the expected gain over the best observed performance $y^*$:
i.e., $EI(\bm x) = \int_{-\infty}^{\infty} \max(y-y^*, 0)p_{M}(y|\bm x)\mathrm{d}y$. 

\subsection{Hyperband Framework}
\label{sec:background_hyperband}

Hyperband (HB)~\cite{li2018hyperband} is a classic MFO framework that halts poorly performing configurations early. 
It consists of three key components: a fidelity partitioning strategy, and nested outer and inner loops responsible for scheduling multi-fidelity evaluations.

\textbf{Fidelity partition} determines how to allocate $r(r\leq R)$ units of resources to evaluate a configuration when the maximum resource is $R$.
As existing methods are particularly applied in AutoML, the typical approaches involve either training for fewer epochs~\cite{jiang2024efficient,li2022hyper,falkner2018bohb} or using a subset of the dataset~\cite{li2021mfes, klein2017fast, li2022hyper}.

\textbf{Inner loop: successive halving (SH)}. 
Given a type of resource partition (e.g., number of epochs or dataset size), HB initially evaluates $n_1$ configurations with each using $r_1$ units of resources, ranking them by performance. HB then continues with only the top $\eta^{-1}$ configurations using $\eta$ times more resources (typically $\eta = 3$), i.e., $n_2 = n_1 \times \eta^{-1}$ and $r_2 = r_1 \times \eta$. This process repeats until the maximum training resource $R$ is reached, i.e., $r_i = R$. 

\textbf{Outer loop: grid search of $n_1$ and $r_1$}. 
HB performs a grid search over $(n_1,r_1)$ pairs to cover different SH brackets. 
Specifically, HB enumerates a bracket index $s\in\{s_{\max},s_{\max}-1,\ldots,0\}$, where $s_{\max}=\lfloor \log_{\eta} R \rfloor$.
Each $s$ defines one feasible pair:
  \[
  n_1=\left\lceil \frac{\mathcal{B}}{R}\frac{\eta^s}{s+1}\right\rceil,
  \qquad
  r_1=R\eta^{-s},
  \]
where $\mathcal{B}=(s_{\max}+1)R$.
Thus, a larger $s$ corresponds to more configurations with fewer initial resources, and vice versa.

%% file: tex/overview.tex
\section{System Overview} 
\label{sec:overview}
Figure~\ref{fig:overview} provides an overview of \sys, which consists of five components:
1) The controller interacts with users and the Spark SQL server, and controls the tuning process.
2) The knowledge database stores both observations and meta-features from the current task as well as historical tasks.
3) The search space optimizer identifies a promising subspace from the original search space.
4) The configuration generator suggests a set of promising configurations.
5) The multi-fidelity optimizer constructs fidelity levels and schedules evaluations across them.

\textbf{Workflow}.
To initiate a tuning task, the user first specifies the tuning objective, the tuning budget, and uploads the target workload to the controller. 
An iterative tuning workflow is then activated:
\textcircled{1} The \textbf{knowledge database} first estimates source-target task similarity using meta-features and historical observations, and provides similar source tasks to other components.
\textcircled{2} The \textbf{search space optimizer} then compresses the original search space using observations from similar tasks, including both knob selection and range narrowing.
\textcircled{3} After that, the \textbf{configuration generator} proposes a batch of promising configurations from the compressed subspace.
Among these, a subset is drawn from top-performing configurations of similar tasks for warm-start, while the rest is generated by Bayesian optimization using weighted surrogates from source tasks and low-fidelity observations.
\textcircled{4} Upon receiving the candidate configurations, the \textbf{multi-fidelity optimizer} iteratively evaluates and prunes them.
If no fidelity partition has been built, it first derives one from similar tasks; this step is performed only once.
It then starts from the lowest fidelity, progressively prunes poor configurations, and increases fidelity until reaching the full workload.
\textcircled{5} Finally, the \textbf{\sys controller} collects the iteration results and stores them in the knowledge database.
These five steps repeat until the user-specified budget, such as time or iteration limits, is exhausted.
After that, \sys selects the best configuration observed in the current task and returns it to the user.

\revise{
  The following sections follow this workflow: 
  Section~\ref{sec:similarity_identification} describes step~\textcircled{1} for identifying transferable source tasks; 
  Section~\ref{sec:search_space_compression} describes step~\textcircled{2} for configuration-space compression; 
  and Section~\ref{sec:multi_fidelity_optimization} describes steps~\textcircled{3}--\textcircled{4}, including candidate generation, query-based fidelity partitioning, and multi-fidelity optimization.
}

\input{tex/method_similarity_identify}

%% file: tex/method_similarity_identify.tex
\section{Similarity Identification}
\label{sec:similarity_identification}
\revise{
    This section describes how \sys identifies historical tasks that are useful for the current tuning task.
}

A continuously deployed Spark SQL tuning system can accumulate historical tasks over time as it performs tuning tasks on different workloads. 
These previous tasks can offer valuable insights for the current tuning. 
For example, workloads from different hardwares or data scales can expose critical queries for selecting representative SQL subsets. 
Even different-query workloads can provide useful signals for identifying promising search-space regions~\cite{zhang2023efficient,shen2023rover}.
\revise{
The intuition is that different Spark SQL workloads can share system-level bottlenecks, such as shuffle parallelism, memory pressure, spill behavior, executor resource allocation, and CPU/I/O contention.
  When two workloads expose similar bottlenecks, configurations that affect these bottlenecks tend to preserve similar performance rankings, even if absolute latencies differ.
}
However, as prior studies show, transferring knowledge from dissimilar tasks can lead to negative impacts~\cite{shen2025tune}. 
Hence, for a given target task, \sys first identifies similar source tasks for safe transfer.

Task similarity is commonly measured by whether two tasks induce similar configuration-performance rankings~\cite{zhang2021restune,sun2025rabbit}.
Given the observations $\mathcal{D}_T$ of the target task, a source task is considered similar if its surrogate can correctly predict the relative performance of $\mathcal{D}_T$, indicating similar system bottlenecks between the two tasks.
Therefore, we define the similarity $S(i,T)$ between the $i$-th source task and the target task using the Kendall-tau coefficient:
\begin{equation}
    S(i,T) = \text{Kendall}^{\mathcal{D}_T} (M_i, Y)
\label{equ:kendall}
\end{equation}
where $M_i$ is the BO surrogate model trained on source observations $\mathcal{D}_i$, 
and $\text{Kendall}^{\mathcal{D}_T}(M_i, Y)$ is the Kendall-tau coefficient between the predictive values and the observed performance on $\mathcal{D}_T$.

\textbf{Warm-starting through prediction}.
However, existing works often overlook a limitation of this similarity metric: 
when $\mathcal{D}_T$ is small, the Kendall-tau estimate in Equation~\ref{equ:kendall} becomes unreliable, 
causing inaccurate and unstable similarity estimates in early tuning iterations.
To mitigate this issue, we extend Rover~\cite{shen2023rover} to warm-start similarity estimation with a similarity predictor.
We first extract query-level features from SparkEventLog under the default configuration and average them as the task meta-feature.
A LightGBM regressor is trained to predict pairwise task similarity from two task meta-features.
It is trained on tuning histories in the knowledge database.
For tasks $i$ and $j$, the label is $\text{Kendall}^{\mathcal{D}_{rand}}(M_i,M_j)$, the Kendall-tau coefficient between their surrogate predictions on random configurations.
After that, for a new tuning task, the trained predictor provides warm-start similarities to all source tasks.

\textbf{Transition mechanism}. 
\revise{
As for the transition from predicted similarity to Kendall-tau similarity computed from target observations, the key insight is that directly computing Equation~\ref{equ:kendall} from only a few target observations can be unstable and may assign misleading weights to source tasks.
  Therefore, \sys keeps using meta-feature-based predicted similarity in the early stage.
  At each iteration, \sys also attempts to compute the Kendall-tau similarity in Equation~\ref{equ:kendall} from the accumulated target observations, but switches to it only when the majority of source tasks have Kendall-tau $p$-values below 0.05.
  Satisfying this condition indicates that the accumulated target observations are sufficient for reliable similarity computation, making it less sensitive to small-sample fluctuations and closer to the true
  source-target relationship.
}

\textbf{Weighting based on similarity}.
After obtaining the similarities of all source tasks, we first filter out source tasks with negative similarity, as they provide less utility for the target task than a random guess. 
Subsequently, the similarities are normalized as transfer weights, where the weight of source task $i$ is $w_i = S(i,T)/\sum_j S(j,T)$.
For generality, \sys also assigns a weight to the target tuning task.
To achieve that, we measure its generalization ability on unseen configurations using the out-of-sample Kendall-tau coefficient.
Specifically, we perform cross-validation on $\mathcal{D}_T$, training the target surrogate on the remaining folds to predict each held-out fold. 
After the transition mechanism enables Equation~\ref{equ:kendall}, the target task also receives a weight based on it.
\revise{
Overall, this weighting mechanism supports all transfer-related components in \sys :
  It aggregates promising regions for search-space compression (Section~\ref{sec:search_space_compression}), guides SQL-subset selection
  for fidelity partitioning (Section~\ref{sec:fidelity_partition}), and combines surrogates for candidate generation (Section~\ref{sec:mfo_process}) 
}

%% file: tex/method.tex
\input{tex/method_cs_compression}

\input{tex/method_mfo}

%% file: tex/method_cs_compression.tex
\section{Configuration Space Compression}
\label{sec:search_space_compression}
To handle the huge Spark configuration space, it is necessary to compress the original space into a promising subspace to improve exploration efficiency~\cite{zhang2023efficient,wei2025toptune,kanellis2022llamatune}.
  Search-space compression includes knob selection and value-range narrowing.
  For knob selection, prior methods use sensitivity analysis, such as Lasso~\cite{tibshirani1996regression} and SHAP~\cite{lundberg2017unified}, or low-dimensional projections, such as HeSBO~\cite{letham2020re} and ALEBO~\cite{nayebi2019framework}
  ~\cite{xin2022locat,fekry2020tune,li2023towards,kanellis2022llamatune,wei2025toptune}.
  For value-range compression, a common strategy is to transfer promising regions from similar tasks~\cite{zhang2023efficient,sun2025rabbit}.
  However, they often define ranges only by the boundaries of discrete historical configurations, which is sensitive to outliers and likely to miss promising but underexplored values.
\revise{
This section explains how \sys compresses the search space by first extracting promising regions from each source task and then aggregating them into selected knobs and value ranges.
}

\subsection{Promising Regions Extraction}
\label{sec:promising_regions_extraction}
\sys first extracts the promising discretized region for each knob from similar source tasks.
Specifically, for source task $i$ with observations $\mathcal{D}_i$, 
we calculate the median performance of all recorded configurations, denoted as $f^i_\text{median}$.
The promising set $G_i$ contains configurations better than this median:
i.e., $G_i = \{\bm{x}\in \mathcal{D}_i \mid f^i(\bm{x}) < f^i_\text{median} \}$, where $f^i(\bm{x})$ is the performance of $\bm{x}$ in the source task.

Although all configurations in $G_i$ are promising, not every knob value in these configurations necessarily contributes positively to the performance. 
Therefore, we employ SHAP to analyze the importance of each knob. 
\revise{
  We choose SHAP because it provides local, value-level, and sign-sensitive attribution, which allows \sys to identify knob values that reduce latency.
  In contrast, Lasso, Gini importance, and fANOVA mainly provide global importance or scores, rather than signed contributions of specific knob values.
}
Specifically, SHAP estimates the contribution of each knob value in every configuration of $G_i$. 
A positive SHAP value means the knob value increases latency, while a negative value means it reduces latency.
We then define the promising knob values as those with negative SHAP values, assuming latency minimization.
For knob $k_j$, its promising value set $P^i_j$ in source task $i$ is:
\begin{equation}
\begin{split}
    P^i_j = \Big\{&\left(\bm{x}[k_j],\ v(\bm{x}) \right)  \mid \bm{x} \in G_i \text{ and } \text{SHAP}(\bm{x}[k_j]) < 0 \Big\}, \\ 
    v(\bm{x}&) = w_i \cdot \frac{f^i_{\text{median}} - f^i(\bm{x})}{f^i_{\text{median}}},
\end{split}
\label{equ:P^i_j}
\end{equation}
where $\bm{x}[k_j]$ denotes the value of $k_j$ in configuration $\bm{x}$ and $v(\bm{x})$ is the weight of this value.
We define $v(\bm{x})$ as the product of the source task's weight $w_i$ and the relative performance improvement of configuration $\bm{x}$ over $f^i_{\text{median}}$, so values from better configurations in more similar tasks receive larger weights.

\subsection{Density Modeling and Target Search Space}
\label{sec:density_modeling_and_target_space}
After extracting promising discrete regions from each source task, \sys constructs the target search space.
For each knob $k_j$, if the weighted majority of tasks provide no promising values, i.e., $\sum_{i=1}^{K} w_i \cdot \mathbb{I}(P^i_j = \emptyset) > 0.5$, \sys removes this generally unpromising knob from the search space.
Otherwise, \sys constructs the final promising value set $P_j$ for $k_j$ by merging the promising values, i.e., 
$P_j = \bigcup_{i=1}^{K} P^i_j$.
Instead of directly using the boundaries of the discrete historical values, \sys estimates their empirical density to find regions where promising values concentrate.
This smooths the promising-value evidence transferred from similar tasks.



For continuous knobs, we use \textbf{Kernel Density Estimation (KDE)}~\cite{parzen1962estimation} to estimate the empirical density of their promising values.
  Specifically, for each continuous knob, we collect the values corresponding to promising configurations and use these samples, along with their associated weights, to construct a weighted density curve. The weighted KDE function is defined as:
\begin{equation}
\hat{g}(x) = \frac{1}{h \cdot \sum_{(\theta, v) \in P_j} v} \sum_{(\theta, v) \in P_j} v \cdot K\left(\frac{x - \theta}{h}\right),
\label{equ:kde_continuous}
\end{equation}
where each $(\theta, v)$ pair represents a promising value and its weight, $h$ is the bandwidth determined by Silverman's rule of thumb, and $K(\cdot)$ is the kernel function, typically Gaussian.
\revise{We use Silverman's rule as a lightweight default for setting the bandwidth; the resulting weighted sum of local kernels can reflect multiple concentrated regions in the empirical promising-value distribution.}
The weights make values from more promising configurations contribute more to the density curve, providing a smoother and more data-driven representation than the discrete boundaries used in~\cite{zhang2023efficient,sun2025rabbit}.

To compress the range of a continuous knob, \sys selects the densest region from the estimated density $\hat{g}(x)$. 
We define a cumulative density threshold $\alpha\in[0,1]$ as the minimum proportion of total probability mass to retain.
\revise{
  This threshold controls the trade-off between compression strength and conservativeness: 
  a smaller $\alpha$ retains keeps only the densest core regions, while a larger $\alpha$ preserves more candidate regions to reduce the risk of over-pruning.}
Assume the original range of $k_j$ is $R_j\subset\mathbb{R}$, the goal is to identify the smallest sub-region $R_j^{\text{opt}}$ that covers at least $\alpha$ probability mass:
\begin{equation}
    \min_{R_j^{\text{opt}} \subset R_j} \quad |R_j^{\text{opt}}| \quad \quad 
    \text{s.t.}\; \int_{R_j^{\text{opt}}} \hat{g}(x) \, dx \geq \alpha,
\end{equation}
In practice, \sys sorts density values in descending order and accumulates them until the retained probability mass reaches at least the minimum proportion $\alpha$. 
The corresponding values define the compressed range $R_j^{\text{opt}}$.

To process categorical knobs, we redefine Equation~\ref{equ:kde_continuous} in a discrete form to build the density function with the indicator function $\mathbb{I}(\cdot)$:
\begin{equation}
    \hat{g}(x) = \frac{1}{\sum_{(\theta, v) \in P_j} v} \sum_{(\theta, v) \in P_j} v \cdot \mathbb{I}(x = \theta).
\end{equation}
The remaining steps follow the continuous-knob case, yielding a unified framework for categorical and continuous knobs.

Compared to existing methods, our approach offers three advantages:
1) It does not require manually setting the number of retained knobs. Through aggregating promising regions from similar tasks, it performs knob selection and range compression jointly.
2) Density modeling captures concentrated promising regions, making compression less sensitive to outliers and able to retain nearby underexplored values.
3) As the similarity estimate of source tasks become more accurate during tuning, the compressed search space can be updated dynamically.

%% file: tex/method_mfo.tex
\section{Multi-Fidelity Optimization}
\label{sec:multi_fidelity_optimization}
\revise{
Following similarity identification and space compression, 
 \sys \sys applies MFO within the subspace.
Specifically, this section introduces how \sys 
 1) partitions fidelity levels, 
 2) generates candidate configurations, 
 and 3) executes the MFO process.
}

\subsection{Fidelity Partition}
\label{sec:fidelity_partition}
As discussed in Section~\ref{sec:background_hyperband}, fidelity partitioning is central to MFO.
We treat evaluation on the original workload as full fidelity, e.g., running 99 queries on 1TB of data.
A $\delta$-fidelity evaluation ($0<\delta\leq1$) uses a $\delta$ fraction of the full-fidelity resource cost, such as execution time or memory consumption. 
For effective pruning in MFO, low-fidelity evaluations should preserve the relative ranking of configurations, so configurations that perform well at full fidelity are not mistakenly discarded.
Since existing low-fidelity choices, such as smaller datasets, early stopping, and database what-if interfaces, are unreliable or unavailable for Spark SQL tuning, \sys constructs Spark SQL-specific fidelities by selecting query subsets.

Assume that the original workload consists of a set of queries $Q=\{q_1,q_2,\dots,q_m\}$.
  Fidelity partitioning aims to select a representative subset $Q_{\delta}\subseteq Q$ that preserves the full-workload evaluation ranking while reducing evaluation cost.
Specifically, for $\delta$-fidelity evaluation, the selected subset $Q_{\delta}$ should satisfy a cost constraint $\text{Cost}(Q_{\delta}) \leq \delta \cdot \text{Cost}(Q)$. 
The problem is formulated as:
\begin{equation}
    \max_{Q_{\delta} \subseteq Q} \ \tau(Q_{\delta}, Q) \quad \text{s.t.} \quad \text{Cost}(Q_{\delta}) \leq \delta \cdot \text{Cost}(Q),
\label{equ:partition_problem}
\end{equation}
where $\tau(Q_{\delta}, Q)$ measures the correlation between the subset $Q_{\delta}$ and $Q$. 
Given per-query performance across configurations, we compute $\tau$ using rank-based metrics such as Kendall-tau.

However, directly estimating $\tau$ on the target task requires sufficient full-fidelity evaluations, which is prohibitively expensive.
To address this, \sys transfers evidence from source tasks with the same query workload but different execution conditions. 
\revise{
Such same-query source tasks naturally arise in practice, including:
1) \textit{data-scale evolution}, 
where the same query set is retuned as data grows or is refreshed; 
2) \textit{hardware/resource migration}, 
where the workload moves across clusters, instance types, or CPU/memory settings; 
and 3) \textit{system performance testing}, 
where teams repeatedly run the same analytical workload for regression testing, benchmarking, or capacity planning. 
In these scenarios, the query set remains stable while data scale, hardware, Spark/runtime version, or resource environment changes.
Therefore, these source tasks can provide evidence for selecting representative SQL subsets without extensive new full-fidelity evaluations on the target task.
}
Moreover, transferring critical-query evidence from similar source tasks is also reasonable. 
Task similarity implies that configurations have consistent relative performance across tasks, indicating shared system bottlenecks.
This consistency suggests that queries exposing key bottlenecks in one task are likely important in another.
Therefore, we first select source tasks with the same query set as the target task. 
For the $i$-th source task, let $\mathcal{D}_i = \{(\bm{x}, \mathbf{p}_{\bm{x}}, \mathbf{c}_{\bm{x}})\}_{\bm{x} \in \mathcal{C}_i}$ denote its observations, where $\mathcal{C}_i$ is the evaluated configuration set.
Here, $\mathbf{p}_{\bm{x}} = [p_{q_1, \bm{x}}, p_{q_2, \bm{x}}, \dots, p_{q_m, \bm{x}}]$ records the per-query performance of configuration $\bm{x}$, 
and $\mathbf{c}_{\bm{x}} = [c_{q_1, \bm{x}}, c_{q_2, \bm{x}}, \dots, c_{q_m, \bm{x}}]$ records the corresponding per-query costs. 
Commonly, an aggregation function $\text{Agg}(\cdot)$ combines per-query performance into workload-level performance for a query set.
For example, the latency metric can be aggregated by summation, i.e., $\text{Agg}(Q_{\text{var}}, \mathbf{p}_{\bm{x}}) = \sum_{q \in Q_{\text{var}}} p_{q, \bm{x}}$, where $Q_{\text{var}}$ is a query set. 
The correlation score of $Q_{\delta}$ in source task $i$ is:
\begin{equation}
\begin{split}
    \tau_i(Q_{\delta}, Q) &= \text{KendallTau}(A^{Q_{\delta}}_i ,\; A^{Q}_i), \\
    A^{Q_{\text{var}}}_i &= \big\{\text{Agg}(Q_{\text{var}}, \mathbf{p}_{\bm{x}}) \mid \bm{x} \in \mathcal{C}_i\big\}.
\end{split}
\label{equ:query_subset_score}
\end{equation}
After that, the overall correlation score is defined as the weighted sum of all source task scores, i.e., $\tau(Q_{\delta}, Q)=\sum_i w_i\tau_i(Q_{\delta}, Q)$.

\begin{algorithm}[t]
\caption{Greedy Solution for Query Subset Selection}
\label{alg:greedy_solution}
\begin{algorithmic}[1]
\REQUIRE Full query set $Q$, fidelity $\delta$, source observations $\{\mathcal{D}_i\}_{i=1}^{K'}$
\ENSURE Optimal subset $Q_{\delta}$
\STATE Initialize $Q_{\delta} \gets \emptyset$ and cost ratio $r\gets0$ \COMMENT{Start with empty}
\STATE Compute the weighted average cost fraction of each query: \\ $c(q) = \sum_{i} w_i \cdot \frac{\sum_{\bm{x} \in \mathcal{C}_i} c_{q, \bm{x}}}{\sum_{q_j \in Q} \sum_{\bm{x} \in \mathcal{C}_i} c_{q_j, \bm{x}}}$.
\WHILE{True}
     \STATE $q^* \gets \arg\max_{q \in Q \setminus Q_{\delta},\ r+c(q)\leq\delta}  \; \tau(Q_{\delta} \cup \{q\}, Q)$
    \STATE \textbf{if} \; $q^*$ is null \; \textbf{then break}
    \STATE $Q_{\delta} \gets Q_{\delta} \cup \{q^*\}$, $r \gets r + c(q^*)$
\ENDWHILE
\RETURN $Q_{\delta}$
\end{algorithmic}
\end{algorithm}

\textbf{Greedy solution}. We observe that Problem~\ref{equ:partition_problem} has transformed into a combinatorial optimization problem. 
We solve it approximately with a greedy algorithm. 
It starts with an empty set and iteratively adds the query that maximizes the correlation score, until the average cost of $Q_{\delta}$ exceeds $\delta$ fraction of the full query set. 
Algorithm~\ref{alg:greedy_solution} presents the pseudocode of the algorithm.

\subsection{Efficient Candidate Generation}
Each HB-based MFO inner loop starts with a set of candidate configurations.
This section describes how \sys recommends them.

Inspired by BOHB~\cite{falkner2018bohb}, we use BO~\cite{hutter2011sequential} for configuration recommendation, as it balances exploration and exploitation better than Hyperband's random sampling. 
However, BO is often inaccurate in early iterations due to sparse observations.
To mitigate this 'cold-start' issue, prior works integrate surrogate models from historical tasks to guide the search~\cite{zhang2021restune,li2023towards}.
In the scenario of MFO, MFES~\cite{li2021mfes} further builds and combines surrogates for observations at different fidelity levels. 
\sys combines these two ideas by treating both historical tasks and fidelity-specific observations as source tasks. 
It assigns their weights using Section~\ref{sec:similarity_identification} and combines their surrogates accordingly. 
To avoid scale mismatch across surrogate outputs, \sys aggregates acquisition-score rankings rather than raw scores.
For sampled and mutated candidates, let $R_i(\bm{x})$ be the rank of configuration $\bm{x}$ under source-task surrogate $M_i$.
The combined rank is $R(\bm{x})=\sum_i w_iR_i(\bm{x})$, and \sys recommends the top-$n_1$ configurations required by the MFO outer loop.

\textbf{Two-phase warm start}. 
To further improve the tuning efficiency, many works warm-start with configurations selected from similar tasks~\cite{li2023towards,shen2025tune,li2025loftune}. 
However, warm start faces an exploration-exploitation trade-off: using too few historical configurations underuses transfer knowledge, while using too many consumes the exploration budget for the target task. 
To address this, \sys designs a two-phase warm-start strategy tailored to MFO.
Phase 1 is similar to existing methods. 
\sys selects the best configurations from the most similar historical tasks and evaluates them at full fidelity at the beginning of tuning. 
After this, \sys starts MFO and transitions to the Phase 2 warm start. 
Specifically, it collects the remaining configurations that outperform the median in each source task, i.e., $G_i$ in Section~\ref{sec:promising_regions_extraction}, and forms the warm-start set $G_\text{ws} = \bigcup_{i=1}^{K} G_i$. 
Configurations in $G_{\text{ws}}$ are ranked by $v(\cdot)$ in Equation~\ref{equ:P^i_j}, so better configurations from more similar tasks receive higher priority. 
At the start of each inner loop of MFO, except loops only containing full-fidelity evaluations, \sys recommends several configurations from $G_{\text{ws}}$ and uses BO to fill the remaining positions up to $n_1$.
The number of configurations drawn from $G_\text{ws}$ is set to the number that will reach full-fidelity evaluation in each inner loop, ensuring that promising configurations do not prune each other. 
Compared with existing strategies, warm start in \sys has two strengths: 
1) It adopts two levels of utilization for configurations with varying priorities. 
2) It balances between exploration and exploitation by making fuller use of historical observations without hindering exploration. 
As most candidates remain BO-generated, and weak Phase-2 warm-start configurations will be filtered out at low-fidelity before wasting excessive resources.

\subsection{Multi-Fidelity Optimization Process}
\label{sec:mfo_process}

Combining the above strategies enables \sys to perform MFO on the compressed search space.
  Before MFO starts, \sys runs Algorithm~\ref{alg:greedy_solution} once to construct query subsets for different fidelity levels.
  The configuration generator then uses warm start and the combined surrogate to recommend candidate configurations.
Finally, \sys follows the Hyperband framework for MFO.

\textbf{Tuning across diverse historical task scenarios}.
\revise{
\sys adapts to different levels of historical-data availability. We consider three scenarios.
1) \textit{Same-query history is available}: 
\sys activates fidelity partitioning, transfer learning, and search-space compression from the beginning with weighted source tasks.
2) \textit{No historical task has the same query set}: 
fidelity partitioning cannot be activated immediately because Equation~\ref{equ:query_subset_score} requires query correspondence.
Thus, the target workload is evaluated at full fidelity in the initial phase.
Meanwhile, transfer learning and search-space compression are still enabled from the beginning, because they rely on task-level similarity and promising configuration regions rather than identical query sets.
As more target observations are gathered, the transition mechanism in Section~\ref{sec:similarity_identification} determines when the current task has accumulated sufficient evidence to serve as a source task for fidelity partitioning;
then \sys constructs low-fidelity query subsets using Equation~\ref{equ:query_subset_score} (e.g., $\tau_T(Q_{\delta},Q)$) and Algorithm~\ref{alg:greedy_solution}, and activates MFO.  
3) \textit{No historical data is available}:  
\sys initially falls back to vanilla BO. 
As target observations accumulate, the transition mechanism jointly enables search-space compression and MFO online.
To sum up, \sys does not permanently depend on same-query historical tasks; its components are activated either from the beginning or online depending on the available evidence.
}

%% file: tex/experiment.tex
\section{EXPERIMENTAL EVALUATION}

\subsection{Experiment Setup}
\label{sec:experiment_setup}

\textbf{Baselines}.
We compare \sys with six baselines. 
--- \textit{Three approaches without transfer learning}: 
\textit{1) LOCAT} \cite{xin2022locat} gradually identifies the important knobs to reduce the search space. 
\textit{2) TopTune} \cite{wei2025toptune}, a DBMS tuning method that compresses the search space with knob-dimensional projection and alternating between categorical and continuous knobs. 
\revise{\textit{3) BOHB}~\cite{falkner2018bohb}, a generic MFO method. We adapt it to Spark SQL with data-volume reduction as the low-fidelity proxy.}
--- \textit{Three approaches with transfer learning}: 
\textit{4) Tuneful} \cite{fekry2020tune} explores significant knobs and applies a multi-task GP with similar source tasks. 
\textit{5) Rover} \cite{shen2023rover} weights surrogates of source tasks by similarity. 
\textit{6) LOFTune} \cite{li2025loftune} warm-starts from similar workloads and uses a workload-aware simulator to fit all historical data. 

\textbf{Benchmarks}. 
We mainly evaluate \sys on two typical benchmarks following literature~\cite{alipourfard2017cherrypick, fekry2020tune, li2025loftune}. 
\textit{1) TPC-H}, a decision support benchmark with 22 queries designed to simulate business-oriented ad-hoc queries and concurrent data modifications.
\textit{2) TPC-DS}, a comprehensive benchmark with 99 queries for evaluating decision support systems.
We utilize two data sizes (100 and 600GB) for both TPC-H and TPC-DS, resulting in four benchmark/data-size combinations.
\revise{
In addition, we use DSB~\cite{ding2021dsb} for the query-shift analysis in Section~\ref{sec:exp_query_shift}, because it supports controlled workload generation.
}

\textbf{Environment}.
We conducted the experiments using Spark 3.4.4 on the OpenEuler 22.03 operating system~\cite{openEuler}.
To simulate tuning and knowledge transfer across different hardware environments, experiments were conducted on eight hardware scenarios, varying in node count (2 or 3), CPU cores (32 or 64), and memory (128GB or 256GB), as shown in Table~\ref{tab:hardware_scenarios}.
Unless otherwise specified, Scenario A serves as the default hardware environment for our evaluations.

\begin{table}[t]
\caption{Hardware Scenarios}
\label{tab:hardware_scenarios}
\centering
\setlength{\tabcolsep}{1.6mm} 
\begin{tabular}{c|cccccccc}
\toprule
\textbf{Scenario} & \textbf{A} & \textbf{B} & \textbf{C} & \textbf{D} & \textbf{E} & \textbf{F} & \textbf{G} & \textbf{H} \\
\midrule
\textbf{Node} & 3 & 3 & 3 & 3 & 2 & 2 & 2 & 2 \\
\textbf{CPU (cores)} & 64 & 32 & 32 & 64 & 64 & 32 & 32 & 64 \\
\textbf{RAM (GB)} & 256 & 128 & 256 & 128 & 256 & 128 & 256 & 128 \\
\bottomrule
\end{tabular}
\end{table}

\textbf{Historical data}.
We collect task meta-features and observation histories of 32 distinct tuning tasks, derived from the combination of benchmarks, data sizes, and eight hardware scenarios. 
Each meta-feature is a 34-d vector extracted from SparkEventLog. 
The history for each task was prepared by running vanilla Bayesian optimization~\cite{hutter2011sequential} and collecting 50 observations. 
By default, we use a leave-one-out evaluation setting, where each target task is tuned using histories from the other 31 tasks.

\textbf{Tuning Setting.}
\textit{1) Optimization target}: In all experiments, we optimize the total latency (i.e., the sum of all query times) of the target Spark SQL workload, and use elapsed time as the cost metric for fidelity partitioning. 
\textit{2) Search space}: We extend the search space in Tuneful~\cite{fekry2020tune} to include 60 parameters that influence the application performance. 
More details are available in the code repository.
\textit{3) Tuning budget}: Except in the cold-start setting, each method runs for 48 hours per target task, and we report the best performance among the configurations it recommends and evaluates. 
\textit{4) Implementation details}: 
\sys implements BO based on OpenBox~\cite{jiang2024openbox},
a toolkit for black-box optimization. 
We set the cumulative density threshold $\alpha=0.65$. 
For MFO, we set the reduction factor $\eta=3$ and the maximum resource $R=9$ following  Section~\ref{sec:background_hyperband}. 
This setting yields two low-fidelity levels, $1/9$ and $1/3$ of the full resource.

\begin{figure*}[tb]
	\centering
	\scalebox{1}[1] {
	\includegraphics[width=0.92\linewidth]{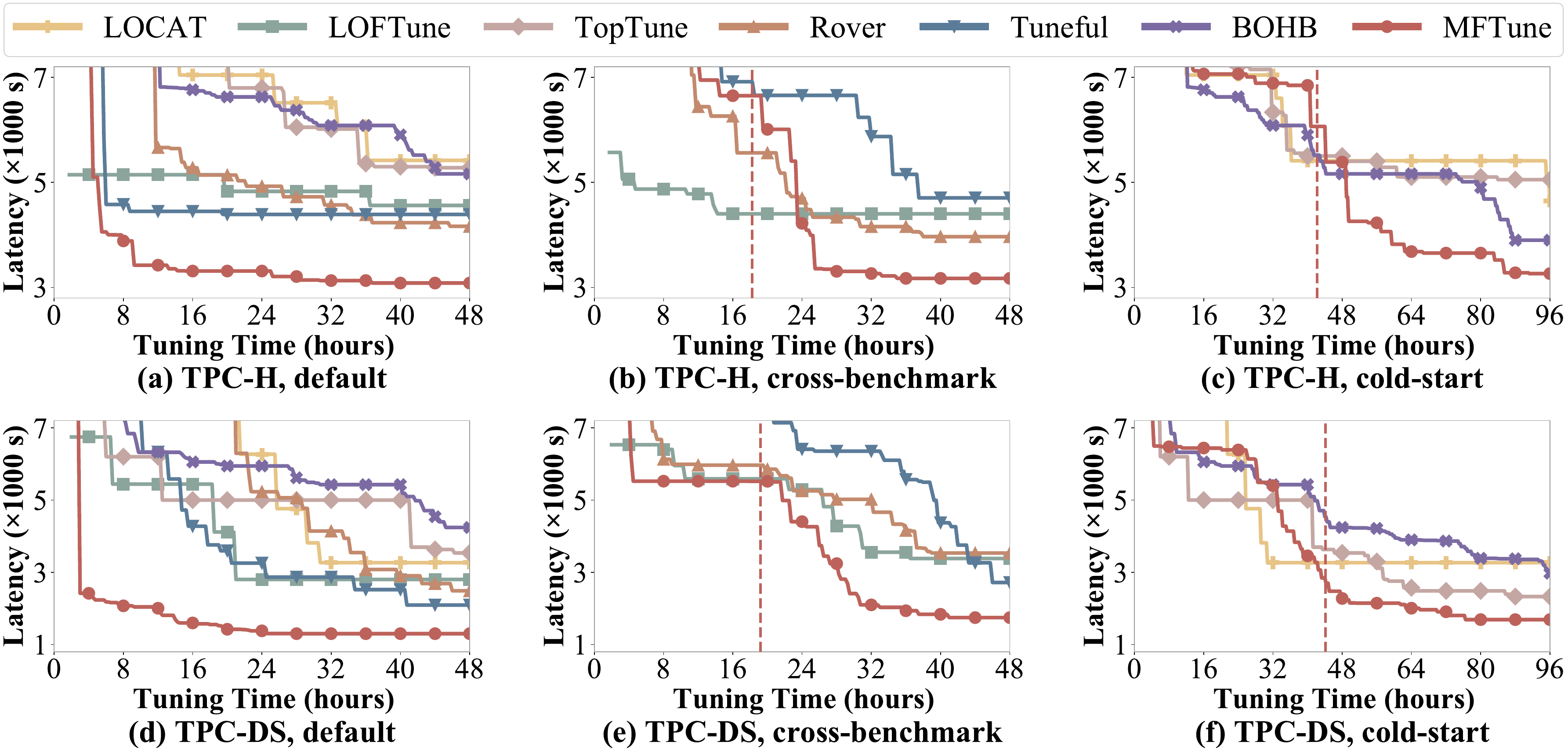}}
	\caption{
    Tuning performance over time across six scenarios. 
    Each curve shows the average best-observed latency over three runs. 
    The vertical red dashed line marks when \sys activates MFO if fidelity partitioning is initially inapplicable.}
    \label{fig:main_exp}
\end{figure*}

\subsection{Main Results under the Default Setting}


This section evaluates \sys on TPC-H and TPC-DS with Hardware Scenario A and 600GB data.
We adopt a leave-one-out pattern, where each target task uses histories from the other 31 tasks. 
Since the repository includes source tasks with the same query set as the target, \sys can leverage historical knowledge to activate fidelity partitioning and MFO from the beginning.

The experimental results are illustrated in Figure~\ref{fig:main_exp}a and ~\ref{fig:main_exp}d, from which we can draw the following conclusions: 
1) methods that leverage historical data (\sys, Tuneful, Rover, and LOFTune) generally exhibit superior performance and faster convergence compared to those starting from scratch (LOCAT, TopTune, and BOHB), demonstrating that historical knowledge from similar tasks can significantly accelerates the search for high-quality Spark SQL configurations.
\revise{Although BOHB uses multi-fidelity evaluation, its data-volume proxy provides less reliable full-workload feedback for Spark SQL, making its performance close to other no-history baselines.}
2) Among the approaches utilizing historical knowledge, \sys shows the strongest convergence. 
While LOFTune initially achieves rapid latency reduction through warm start, its use of history is limited to initialization. 
This limited use of history leads to stagnation, allowing Rover and Tuneful to eventually surpass it. 
Tuneful converges faster than Rover in the early stage because it combines transfer learning with search-space compression, while Rover does not reduce the search space. 
Although these baselines leverage transfer learning to accelerate convergence, their exclusive reliance on full-fidelity evaluations inevitably leads to time-consuming explorations of poor configurations. 
In contrast, \sys employs MFO to filter poor configurations at low fidelity, and further strengthens this process with search-space compression and tailored transfer learning, enabling sustained latency reduction throughout tuning.
3) \sys yields the best final tuning results across both benchmarks. 
Specifically, compared to the six baselines, \sys reduces latency of Spark SQL ranging from 25.9\% to 43.1\% on TPC-H and 37.8\% to 69.3\% on TPC-DS, respectively.

\subsection{Evaluation on Generalization}
This subsection evaluates the generalization of \sys in four settings:
cross-benchmark transfer, cold start, transfers with varying query overlap, and transfers across data sizes and hardwares.

\subsubsection{Results in Cross-Benchmark Transfer}
\label{sec:exp_results_in_cross_benchmark_transfer}
This section evaluates \sys's robustness under cross-benchmark transfer, where source and target workloads come from different benchmarks. 
We compare \sys against the three baselines that utilize historical data (Tuneful, Rover, and LOFTune), which are the top-3 baselines in the default setting.
In the cross-benchmark setting, when TPC-H is the target benchmark, the framework uses 16 source tasks only from TPC-DS, and vice versa. 
As discussed in Section~\ref{sec:mfo_process}, without same-query history, \sys cannot activate fidelity partitioning at the beginning. 
Thus, \sys initially uses transfer learning and search-space compression with full-fidelity evaluations. 
Once sufficient target observations are collected, it dynamically constructs low-fidelity query subsets and activates MFO.

The experimental results are shown in Figure~\ref{fig:main_exp}b and ~\ref{fig:main_exp}e.
The vertical red dashed line marks the specific time point \sys enables MFO once fidelity partitioning becomes applicable with sufficient observations. 
Following this activation, the performance gap between \sys and the baselines widens significantly as low-fidelity proxies enable high-frequency explorations beyond the reach of full-fidelity methods.
Overall, \sys also achieves the best tuning results under cross-benchmark transfer.
Compared with the baselines, \sys achieves a latency reduction ranging from 20.0\% to 32.5\% on TPC-H and 35.7\% to 50.6\% on TPC-DS, showing its adaptability when same-query history is unavailable.

\subsubsection{Results without Historical Data}
\label{sec:exp_without_history}

To further evaluate the robustness of \sys in extreme scenarios, we conduct experiments under a cold-start setting where no historical data is available. In this subsection, we compare \sys against TopTune, LOCAT, and BOHB, the three baselines that do not rely on historical knowledge. Unlike the previous settings, the tuning budget here is extended to 96 hours to allow for sufficient exploration from scratch on the 600GB data scale. As detailed in Section~\ref{sec:mfo_process}, since there are no historical tasks to leverage, \sys initially degrades to a vanilla BO. As the process continues and observations accumulate, the system automatically enables space compression and MFO.

Figure~\ref{fig:main_exp}c and ~\ref{fig:main_exp}f illustrate the experimental results for TPC-H and TPC-DS, respectively. 
During the initial tuning phase, \sys exhibits a similar convergence trend to the baselines while gathering the necessary observations.
However, once MFO is activated (marked by the vertical red dashed line), the convergence speed increases significantly, demonstrating that \sys can successfully identify and utilize low-fidelity proxies online even without prior history.
Eventually, \sys achieves a much lower final latency compared to TopTune and LOCAT, which remain limited by the high temporal cost of performing exclusive full-fidelity evaluations. Specifically, \sys outperforms all no-history baselines by 16.3\% to 35.4\% on TPC-H, and by 27.4\% to 48.2\% on TPC-DS.


\begin{table*}[t]
  \centering
  \setlength{\tabcolsep}{6pt}
  \caption{\revise{Effect of query overlap on DSB. For example, AB$\rightarrow$AC means tuning AC with AB as the source. Latencies are reported in $10^3$ seconds. Early best is measured at 25\% of the tuning budget. Gains are computed relative to \sys without AB history.}}
  \label{tab:workload_shift_overlap}
  \begin{tabular}{@{}lcccccccc@{}}
  \toprule
  \multirow{2}{*}[-2pt]{\textbf{Workload shift}} &
  \multirow{2}{*}[-2pt]{\textbf{Overlap}} &
  \multirow{2}{*}[-2pt]{\textbf{Default latency}} &
  \multicolumn{3}{c}{\textbf{Early best latency}} &
  \multicolumn{3}{c}{\textbf{Final best latency}} \\
  \cmidrule(lr){4-6}
  \cmidrule(lr){7-9}
  &
  &
  &
  \textbf{w/o AB} &
  \textbf{w/ AB} &
  \textbf{Gain} &
  \textbf{w/o AB} &
  \textbf{w/ AB} &
  \textbf{Gain} \\
  \midrule
  AB$\rightarrow$ABC & 71.2\% & 3.22 & 2.79 & 0.98 & 64.9\% & 1.58 & 0.82 & 48.1\% \\
  AB$\rightarrow$AC  & 50.0\% & 1.86 & 1.42 & 1.01 & 28.9\% & 0.71 & 0.48 & 32.7\% \\
  AB$\rightarrow$C   & 0.0\%  & 0.84 & 0.62 & 0.58 & 6.5\%  & 0.26 & 0.22 & 15.4\% \\
  \bottomrule
  \end{tabular}
\end{table*}

\begin{figure*}[!t]
	\centering
	\includegraphics[width=0.97\linewidth]{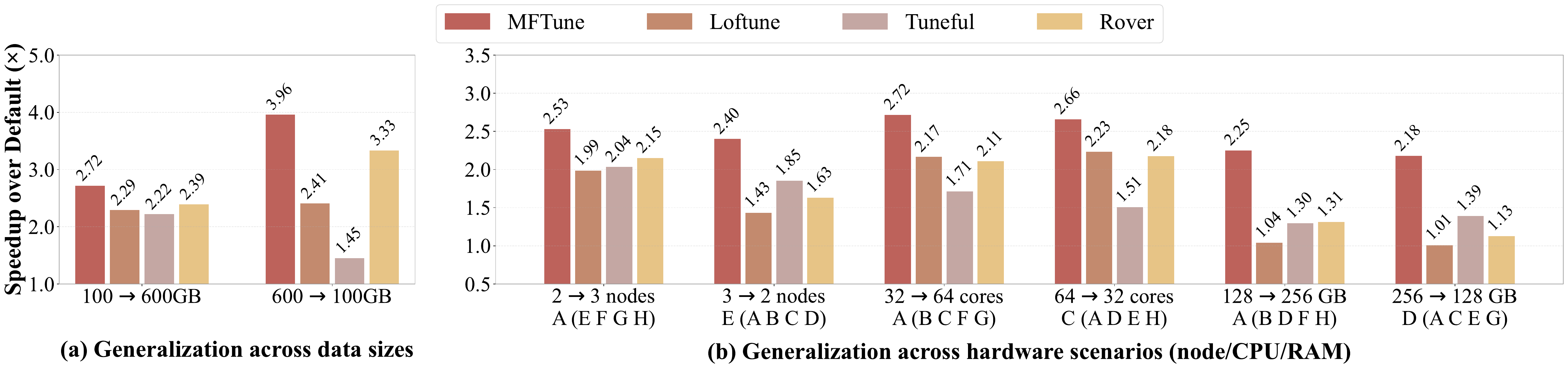}
	\caption{Generalization performance of transfer-learning-based tuning methods across diverse settings. Results represent the speedup of the optimal latency achieved within 48 hours relative to the default Spark configuration.}
    \label{fig:exp_generalization}
\end{figure*}

\subsubsection{\texorpdfstring{\revise{Results under Varying Degrees of Query Shift}}{Results under Varying Degrees of Query Shift}}
\label{sec:exp_query_shift}

\revise{
The default and cross-benchmark settings represent two endpoints of history availability: the former has same-query history, while the latter has no overlap between source and target query sets.
To further examine \sys when no historical task has the same query set, we construct controlled workload shift with varying degrees of query overlap using DSB~\cite{ding2021dsb}.
We choose DSB for two reasons: first, it enhances the data distributions of TPC-DS and provides richer query templates; 
second, it provides explicit query-type classifications for templates, which allows us to construct semantically meaningful workload shifts rather than arbitrary splits.
DSB queries are divided into three groups:
A contains 15 aggregation queries, B contains 22 multi-block queries, and C contains 15 Select-Project-Join queries.
We use AB as the source workload, which contains 37 queries, and evaluate three target workloads with decreasing overlap: ABC (52 queries), AC (30 queries), and C (15 queries).

Table~\ref{tab:workload_shift_overlap} reports the gains of using AB history over no history under different query-overlap levels.
  Although MFO cannot be activated initially, AB history still improves early tuning by guiding transfer learning and search-space compression toward promising configurations and knob regions.
  As overlap decreases from 71.2\% to 50.0\% and 0.0\%, the early gain drops from 64.9\% to 28.9\% and 6.5\%, indicating weaker immediate guidance from less-similar histories.
  The final gain additionally includes the effect of MFO after target observations accumulate, reaching 48.1\%, 32.7\%, and 15.4\% in the three settings.
  These results show a general consistency between query overlap and transfer benefits, indicating that \sys can exploit more useful historical information when source tasks are more related.
  However, consistent with the cross-benchmark transfer, even under complete query drift, historical transfer plus delayed MFO outperforms tuning without history.
}

    
    

\subsubsection{Results with Varying Data Size and Hardware}
\label{sec:exp_varying_data_hardware}

This subsection evaluates \sys on TPC-H under data-size and hardware shifts.
Figure~\ref{fig:exp_generalization}a reports cross-scale transfer between 100GB and 600GB on Hardware A. 
When 600GB is the target, the 16 tasks from the 100GB setting are source tasks, and vice versa. 
Furthermore, Figure~\ref{fig:exp_generalization}b reports transfer across different hardware configurations, including node count, CPU cores, and memory. 
For example, in the $2\xrightarrow[]{}3$ node setting, Hardware A (600GB) is the target, and the 16 tasks from 2-node environments (Hardware E, F, G, H) are source tasks.

The experimental results demonstrate the superior robustness of \sys compared to existing transfer-learning-based methods. 
\sys consistently achieves the highest latency speedup across all tested scenarios. 
Notably, it reaches a peak speedup of 3.96x in the $600\xrightarrow[]{}100$ GB data size transfer. 
In hardware-shift scenarios, while baselines exhibit significant performance volatility, \sys maintains robust speedups exceeding 2.18x. 
\revise{These results show that across different historical transfer scenarios, \sys can adaptively identify and weight similar source tasks through similarity identification and exploit their information to accelerate tuning.}

\subsection{Detailed Analysis}

\begin{figure*}[t]
    \centering
    
    \begin{subfigure}[b]{0.33\linewidth} 
        \centering
        \includegraphics[width=\linewidth]{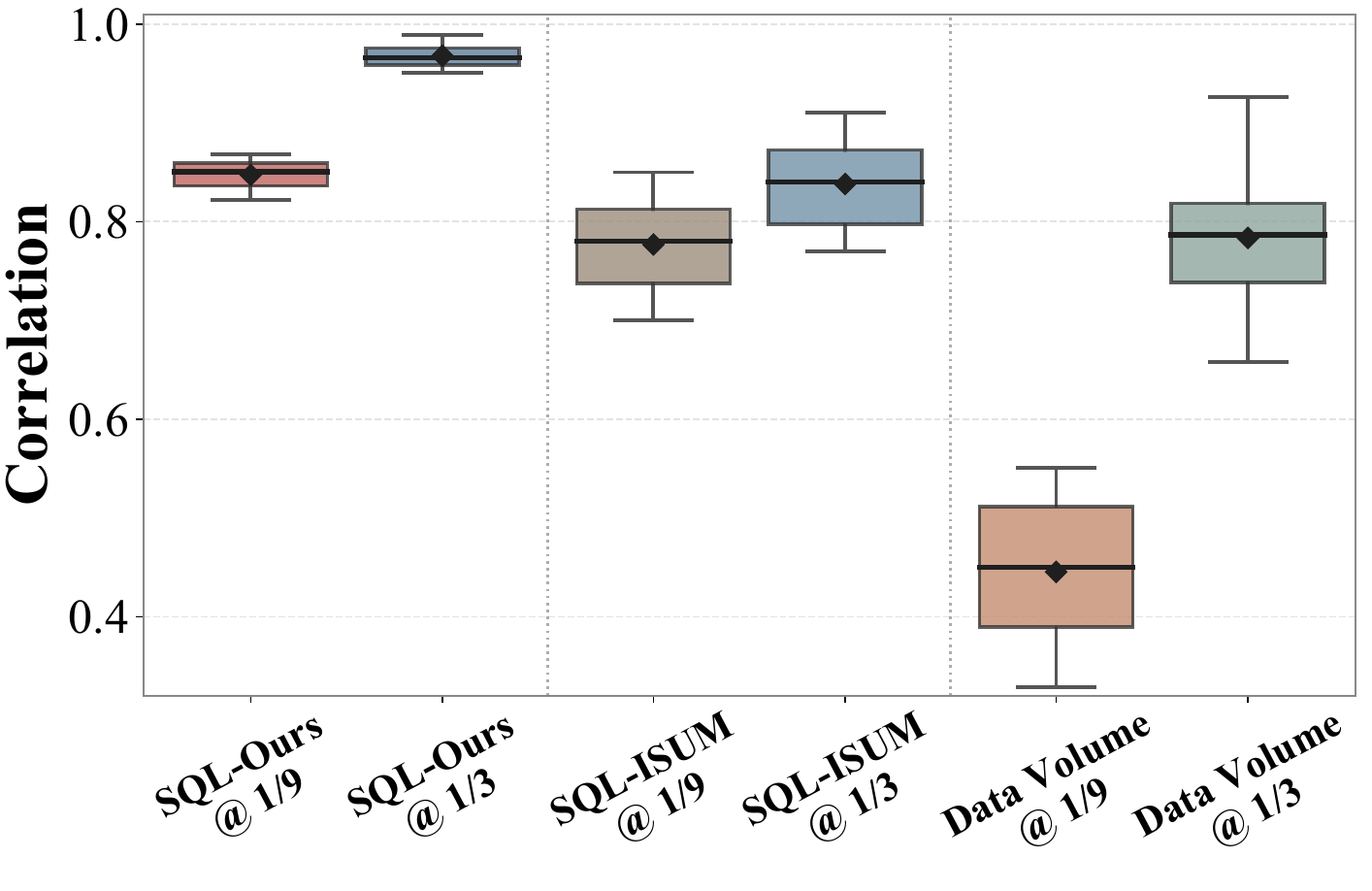}
        \caption{Fidelity correlation.}
        \label{fig:corr_of_fidelity}
    \end{subfigure}
    \hfill
    \begin{subfigure}[b]{0.33\linewidth} 
        \centering
        \includegraphics[width=\linewidth]{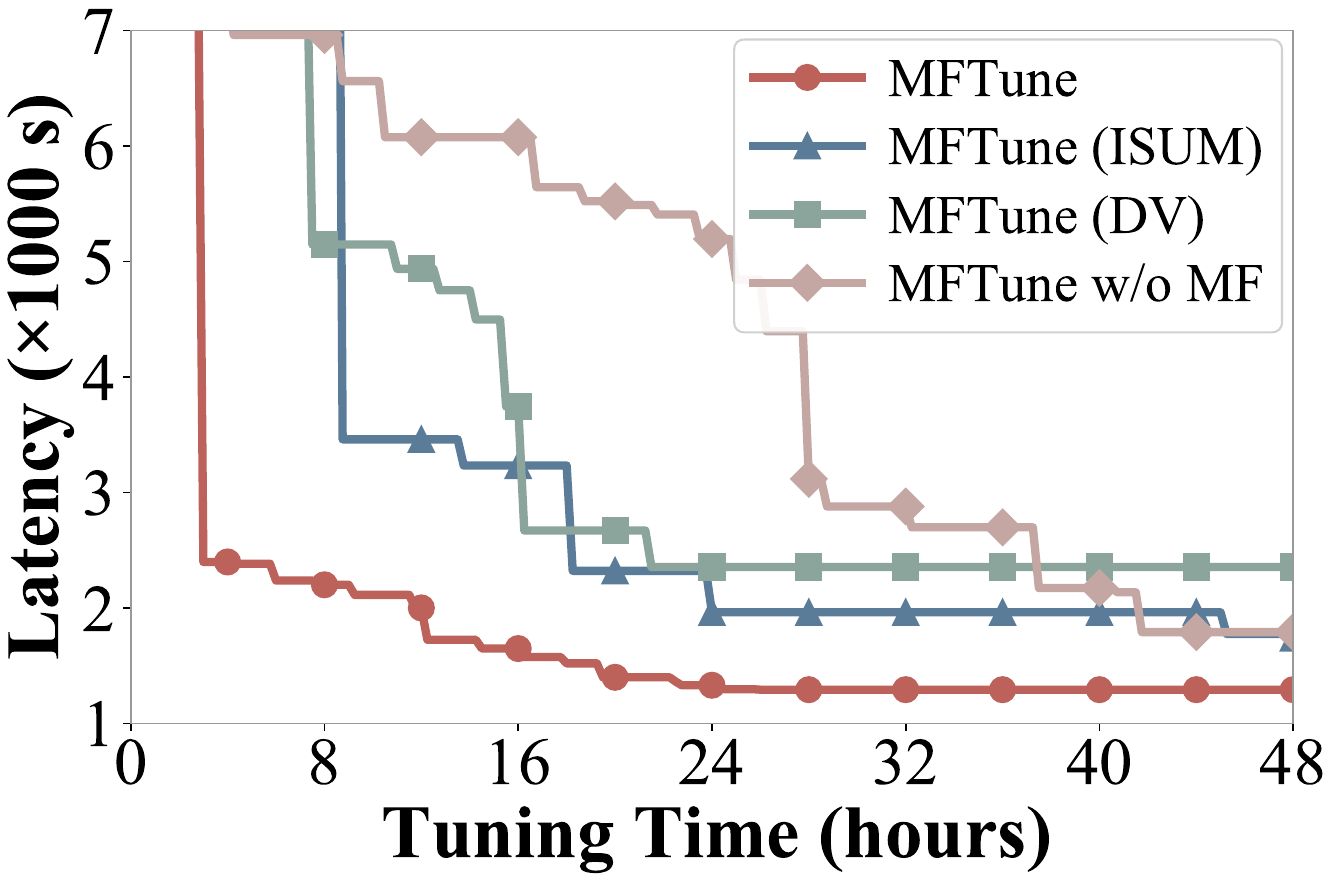}
        \caption{Performance over time.}
        \label{fig:aba_of_mfo}
    \end{subfigure}
    \hfill
    \begin{subfigure}[b]{0.31\linewidth}
        \centering
        \setlength{\tabcolsep}{2.5pt}
        \renewcommand{\arraystretch}{1.1}
        \resizebox{\linewidth}{!}{
        \begin{tabular}{lcc}
        \toprule
            \textbf{\makecell[l] {System\\telemetry}} &
            \textbf{\makecell{$1/9$\\fidelity}} &
            \textbf{\makecell{$1/3$\\fidelity}} \\
        \midrule
        Avg peak memory & 0.86 & 0.92 \\
        GC time & 0.43 & 0.57 \\
        Shuffle read & 0.73 & 0.87 \\
        Spill rate & 0.64 & 0.57 \\
        \bottomrule
        \end{tabular}}
        \vspace{2em}
        \caption{Correlation on system hidden states.}
        \label{fig:telemetry_corr}
    \end{subfigure}
    
    \caption{Effectiveness of the MFO mechanism on TPC-DS.}
    \label{fig:exp_mfo} 
\end{figure*}

\subsubsection{Analysis of Multi-Fidelity Optimization}
\label{sec:exp_analysis_of_MFO}


\revise{
This subsection analyzes the effectiveness of \sys's MFO mechanism on TPC-DS.

We first examine whether existing DBMS workload-compression methods can construct query subsets for Spark SQL MFO.
  These methods cannot be directly applied mainly because their original metrics target DBMS downstream tasks such as index selection and Approximate Query Processing (AQP).
  They rely on signals or tools such as candidate indexes, hypothetical-index what-if interfaces, and AQP-specific error models, which are unavailable in our Spark SQL tuning task.
  Therefore, we keep their subset-selection algorithms where possible, but replace their original signals with transfer-based Spark SQL adaptations.
1) SQL-Dist~\cite{chaudhuri2002compressing} uses application-specific query distances for index selection or AQP.
Since these distances cannot be directly used in Spark SQL tuning, 
we redefine query distance as $1-\tau$, 
where $\tau$ is the Kendall correlation between two queries' source-task latency vectors.
2) SQL-GSUM~\cite{deep2020comprehensive} requires SQL text and execution statistics to preserve feature coverage. We adapt the latter one with SparkEventLog.
3) SQL-ISUM~\cite{siddiqui2022isum} defines utility and influence for index-tuning improvement.
  We adapt the utility by source-task latency improvement over the default configuration and similarity by Kendall correlation between source-task latency vectors.
Overall, these adaptations favor the DBMS workload-compression baselines: SQL-Dist and SQL-ISUM are strengthened with source-task transfer signals from \sys, despite being originally inapplicable to Spark SQL tuning.

Figure~\ref{fig:partition} compares the fidelity quality of these methods across different fidelity levels.
SQL-ISUM and SQL-Dist outperform SQL-GSUM because our adaptations strengthen their utility/similarity or distance metrics with source-task latency correlations.
However, they still underperform SQL-Ours, showing that the mismatch is not only the required signals.
Specifically, their subset-selection algorithms remain misaligned with MFO.
\sys directly optimizes whether the subset preserves the full-workload configuration ranking, while these methods indirectly encourage representativeness through query-to-query distance or influence.
This direct-versus-indirect optimization gap explains why adapted workload-compression baselines still fall behind SQL-Ours.
Overall, SQL-Ours performs better for two reasons: 
1) its metric targets Spark SQL tuning feedback, 
and 2) its greedy selection directly optimizes the consistency with the full-workload required by MFO.
Figure~\ref{fig:corr_of_fidelity} further tests this conclusion across 16 TPC-DS workloads.
  SQL-Ours achieves the highest and most concentrated correlations, with means of 0.847 at $1/9$ and 0.968 at $1/3$, outperforming both SQL-ISUM (0.777 and 0.838) and Data Volume (0.445 and 0.784).

We further analyze why the selected low-fidelity subset correlates well with full-workload latency.
  Figure~\ref{fig:telemetry_corr} reports the Kendall correlation between low-fidelity and full-fidelity telemetry across 50 random configurations.
  We examine four representative Spark SQL hidden-state indicators: peak execution memory, JVM GC time, shuffle read bytes, and spill rate normalized by running time.
  The low-fidelity subset preserves the peak memory and shuffle signals well: correlations reach 0.86/0.92 for peak memory and 0.73/0.87 for shuffle read at $1/9$/$1/3$ fidelity.
  GC time and spill rate are noisier: GC is sensitive to JVM/runtime state, while spill rate can be dominated by a few memory-intensive operators or skewed partitions. 
  Nevertheless, they still show non-trivial correlation.
  These results provide system-level evidence that the subset preserves key bottlenecks, supporting its high correlation with full-workload latency.
  
We finally evaluate whether the fidelity-quality advantage in Figure~\ref{fig:partition} improves end-to-end tuning.
  Specifically, we compare \sys with three variants:
  1) \sys w/o MF, which relies only on full-fidelity evaluations;
  2) \sys (DV), which uses data-volume reduction as the low-fidelity proxy;
  and 3) \sys (ISUM), which replaces our SQL Selection with adapted SQL-ISUM.
  We choose SQL-ISUM because it has the best fidelity correlation among adapted DBMS workload-compression baselines in Figure~\ref{fig:partition}.
Figure~\ref{fig:aba_of_mfo} shows that \sys converges faster and achieves a better final configuration than all variants.
  Compared with \sys w/o MF, \sys (DV), and \sys (ISUM), \sys reduces latency by 27.8\%, 45.1\%, and 25.6\%, respectively. 
  This shows that other fidelity partitioning strategies are insufficient for Spark SQL MFO. 
}


\begin{figure*}[t]
	\centering
	\scalebox{1}[1] {
	\includegraphics[width=0.98\linewidth]{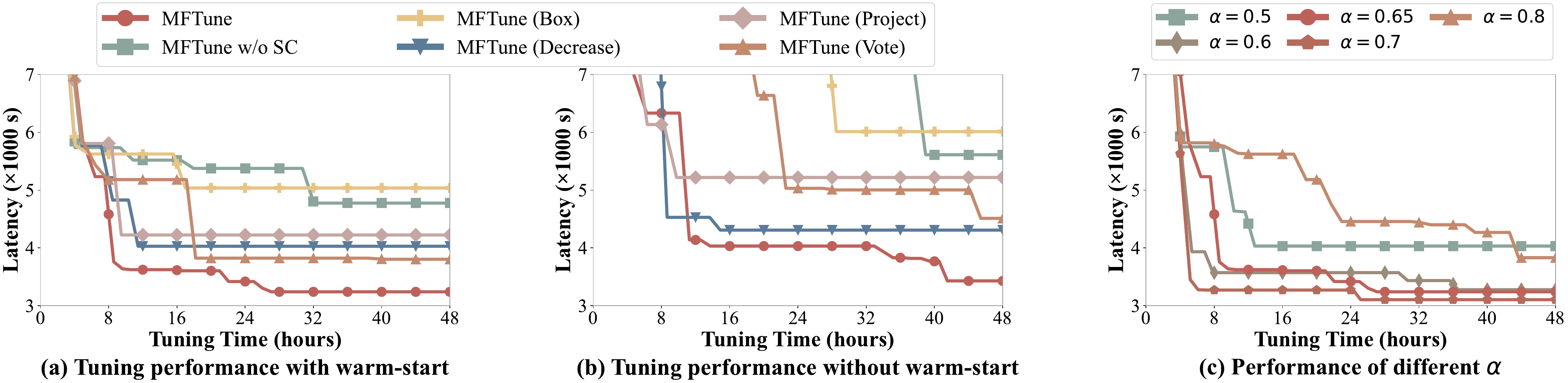}}
	\caption{Analysis of the search space compression (SC) mechanism on TPC-H.}
    \label{fig:exp_sc}
\end{figure*}

\subsubsection{Analysis of Space Compression}
In this section, we evaluate the effectiveness of the search space compression (SC) mechanism in \sys and analyze the sensitivity of hyperparameter $\alpha$.

We first compare \sys’s space compression with the following baselines:
\textit{1) w/o SC}, the uncompressed search space.
\textit{2) Box}~\cite{box_perrone2019learning} suggests a smallest box covering the best-observed configurations of source tasks.
\textit{3) Decrease}, adopted in Tuneful~\cite{fekry2020tune}, which removes 40\% unimportant
knobs every 10 iterations based on the importance ranking. 
\textit{4) Project}, adopted in LlamaTune~\cite{kanellis2022llamatune} and TopTune~\cite{wei2025toptune}, which projects the original space to a synthetic low-dimensional one and narrows ranges by bucketization.
\textit{5) Vote}, adopted in OpAdvisor~\cite{zhang2023efficient}, which extracts boundary values of discrete historical configurations and votes to generate the target space.
Specifically, we replace \sys's SC component with each baseline strategy and run end-to-end comparisons on TPC-H 600GB.

As shown in Figure~\ref{fig:exp_sc}a, 
\sys maintains an advantage throughout tuning, with a final latency reduction of 14.8\%--35.7\% over the variants. 
Existing strategies show clear limitations: Decrease relies on target observations, which are often insufficient for accurate knob ranking; 
Box can over-prune by ignoring task-specific differences; and Project reduces dimensionality but cannot exclude poor subspaces precisely.
Vote is the closest to our approach, its reliance on discrete boundaries makes it highly sensitive to outliers. 
In contrast, \sys employs Kernel Density Estimation (KDE) to smooth out noise and identify robust high-potential regions.
To isolate the effect of SC, we further conduct a stress test by disabling the warm start. 
The rationale is that warm start can hide the weakness of a suboptimal search space by directly evaluating strong historical configurations at the beginning. 
As shown in Figure~\ref{fig:exp_sc}b, without warm start, the gap between \sys and the variants becomes larger, and \sys reduces latency by 20.4\%--43.0\%. 
This demonstrates the robustness of our density-based SC strategy.

Finally, Figure~\ref{fig:exp_sc}c analyzes the sensitivity of the cumulative density threshold $\alpha$.
  This threshold controls the trade-off between pruning strength and conservativeness.
  When $\alpha$ is too small, e.g., 0.5, the space is over-pruned and may exclude the optimum, which causes early stagnation.
  When $\alpha$ is too large, e.g., 0.8, compression is too conservative and retains too many low-potential regions, which may slow the convergence.
  Between these extremes, \sys is stable for $\alpha\in[0.6,0.7]$, achieving comparable convergence.
Therefore, we set $\alpha=0.65$ by default.


\begin{table}[t]
\centering
\caption{Ablation study of warm-start phases on TPC-H (600GB). P1 and P2 denote Phase 1 warm start and Phase 2 warm start, respectively. All metrics represent the relative gains achieved by \sys compared to each specific variant.}
\label{tab:ws_ablation}
\begin{tabular}{@{}cccc@{}}
\toprule
\textbf{P1} & \textbf{P2} & \textbf{Latency Reduction (\%)} & \textbf{Tuning Acceleration} \\ \midrule
$\times$    & $\times$    & 5.50\%                    & 2.15$\times$                \\
$\checkmark$ & $\times$    & 5.13\%                    & 1.98$\times$                \\
$\times$    & $\checkmark$ & 1.25\%                    & 1.12$\times$                \\ \bottomrule
\end{tabular}
\end{table}

\subsubsection{Ablation Study of Warm Start}

This subsection evaluates the contribution of the two-phase warm-start strategy.
  As shown in Table~\ref{tab:ws_ablation}, the two-phase design outperforms both single-phase variants.
  Phase 1 provides a strong initial configuration, while Phase 2 drives further improvement by working with MFO to filter and exploit more candidates at low fidelity.
Overall, the two-phase strategy makes full use of historical data and balances fast initialization with sustained exploration, achieving a 5.50\% latency reduction and 2.15$\times$ tuning acceleration over the cold-start baseline.

\subsubsection{\texorpdfstring{\revise{Robustness to Observational Noise}}{Robustness to Observational Noise}}
  \label{sec:exp_noise_robustness}

\revise{
In real clusters, Spark SQL evaluations can be affected by transient noise from network jitter, CPU contention, or other background activities.
  Thus, we conduct a controlled noise-injection study to test the stability of \sys under increasingly severe noise.
Specifically, for each evaluated latency $y$, we inject multiplicative noise as
  $\tilde{y}=y\cdot\max(1+\epsilon,0.01)$, where $\epsilon\sim\mathcal{N}(0,\sigma^2)$.
  We consider three noise levels: $\sigma=0.05$, $0.10$, and $0.15$.
\begin{table}[t]
  \centering
  \caption{\revise{Noise study on TPC-H (600GB). Best retained $a{\rightarrow}b$ denotes the ratio that the clean best configuration at fidelity $a$ is still promoted to fidelity $b$ after noise injection.}}
  \label{tab:noise_robustness}
  \setlength{\tabcolsep}{3pt}
  \renewcommand{\arraystretch}{1.12}
\resizebox{0.95\linewidth}{!}{%
  \begin{tabular}{lccc}
    \toprule
    \makecell{\textbf{Noise}\\\textbf{level}} &
    \makecell{\textbf{Latency change}\\\textbf{vs. clean}} &
    \makecell{\textbf{Best retained}\\\textbf{$1/9{\rightarrow}1/3$}} &
    \makecell{\textbf{Best retained}\\\textbf{$1/3{\rightarrow}1$}} \\
    \midrule
    $\sigma=0.05$ & -1.23\% & 95.0\% & 78.5\% \\
    $\sigma=0.10$ & 3.17\% & 90.5\% & 70.0\% \\
    $\sigma=0.15$ & 9.64\% & 82.4\% & 55.2\% \\
    \bottomrule
  \end{tabular}%
  }
\end{table}
  Table~\ref{tab:noise_robustness} shows that \sys remains stable under increasing observational noise.
  1) The $1/9{\rightarrow}1/3$ promotion stage has high retention rates, remaining above 82.4\% under the three noise levels.
  This is because this stage promotes multiple configurations, e.g., 9 to 3, giving the best configuration room to survive noisy perturbations.
  2) In contrast, the $1/3{\rightarrow}1$ stage is stricter, often promoting only one configuration from 3 or 5 candidates.
  Therefore, its retention rates drop to 78.5\%, 70.0\%, and 55.2\%.
  3) However, lower second-stage retention does not translate into severe tuning degradation.
  Even under the highest noise, latency increases by only 9.64\%, still smaller than the gaps between \sys and competing baselines in the main experiments.
  This is because configurations reaching $1/3$ fidelity are already promising, so selecting a near-best candidate can still yield competitive final performance.
}

\subsubsection{Analysis of Tuning Overhead}
\label{sec:tuning_overhead}

We record the computational overhead of \sys under the default settings. 
For task-level initialization, source-task similarity prediction takes $\sim$15s. 
Fidelity partitioning takes $\sim$21s for TPC-DS compared to $\sim$0.5s for TPC-H, mainly because TPC-DS has many more queries. 
At the iteration level, similarity calculation via Kendall takes $\sim$0.6s, search-space compression takes $\sim$2s, and BO candidate recommendation takes $\sim$0.2s. 
\revise{
Within search-space compression, SHAP computation takes $\sim$0.3s on average.
}
Collectively, these overheads are negligible compared with configuration evaluations, which take thousands of minutes, and total tuning budgets of dozens of hours.

%% file: sample.bib
@String{Computing = "Computing" }

@String{Computer = "{IEEE} Computer" }

@String{Springer = "Springer-Verlag" }

@inproceedings{armbrust2015spark,
  title={Spark sql: Relational data processing in spark},
  author={Armbrust, Michael and Xin, Reynold S and Lian, Cheng and Huai, Yin and Liu, Davies and Bradley, Joseph K and Meng, Xiangrui and Kaftan, Tomer and Franklin, Michael J and Ghodsi, Ali and others},
  booktitle={Proceedings of the 2015 ACM SIGMOD international conference on management of data},
  pages={1383--1394},
  year={2015}
}

@online{spark_configuration,
  author =       "SparkConf",
  year =         "2025",
  title =        "Configuration - Spark 4.0.1 Documentation.",
  url =          "https://spark.apache.org/docs/latest/configuration.html",
  organization = "Stanford University",
  note =         "Last accessed: July 5, 2026"
}

@inproceedings{snoek2012practical,
  title={Practical bayesian optimization of machine learning algorithms},
  author={Snoek, Jasper and Larochelle, Hugo and Adams, Ryan P},
  booktitle={Advances in neural information processing systems},
  pages={2951--2959},
  year={2012}
}

@inproceedings{hutter2011sequential,
  title={Sequential model-based optimization for general algorithm configuration},
  author={Hutter, Frank and Hoos, Holger H and Leyton-Brown, Kevin},
  booktitle={International Conference on Learning and Intelligent Optimization},
  pages={507--523},
  year={2011},
  organization={Springer}
}

@inproceedings{bergstra2011algorithms,
  title={Algorithms for hyper-parameter optimization},
  author={Bergstra, James S and Bardenet, R{\'e}mi and Bengio, Yoshua and K{\'e}gl, Bal{\'a}zs},
  booktitle={Advances in neural information processing systems},
  pages={2546--2554},
  year={2011}
}

@article{zhang2022facilitating,
  title={Facilitating database tuning with hyper-parameter optimization: a comprehensive experimental evaluation},
  author={Zhang, Xinyi and Chang, Zhuo and Li, Yang and Wu, Hong and Tan, Jian and Li, Feifei and Cui, Bin},
  journal={Proceedings of the VLDB Endowment},
  volume={15},
  number={9},
  pages={1808--1821},
  year={2022},
  publisher={VLDB Endowment}
}

@article{jones1998efficient,
  title={Efficient global optimization of expensive black-box functions},
  author={Jones, Donald R and Schonlau, Matthias and Welch, William J},
  journal={Journal of Global optimization},
  volume={13},
  number={4},
  pages={455--492},
  year={1998},
  publisher={Springer}
}

@inproceedings{falkner2018bohb,
  title={BOHB: Robust and efficient hyperparameter optimization at scale},
  author={Falkner, Stefan and Klein, Aaron and Hutter, Frank},
  booktitle={International conference on machine learning},
  pages={1437--1446},
  year={2018},
  organization={PMLR}
}

@article{li2018hyperband,
  title={Hyperband: A novel bandit-based approach to hyperparameter optimization},
  author={Li, Lisha and Jamieson, Kevin and DeSalvo, Giulia and Rostamizadeh, Afshin and Talwalkar, Ameet},
  journal={Journal of Machine Learning Research},
  volume={18},
  number={185},
  pages={1--52},
  year={2018}
}

@article{li2022hyper,
  title={Hyper-tune: towards efficient hyper-parameter tuning at scale},
  author={Li, Yang and Shen, Yu and Jiang, Huaijun and Zhang, Wentao and Li, Jixiang and Liu, Ji and Zhang, Ce and Cui, Bin},
  journal={Proceedings of the VLDB Endowment},
  volume={15},
  number={6},
  pages={1256--1265},
  year={2022},
  publisher={VLDB Endowment}
}

@inproceedings{jiang2024efficient,
  title={Efficient hyperparameter optimization with adaptive fidelity identification},
  author={Jiang, Jiantong and Wen, Zeyi and Mansoor, Atif and Mian, Ajmal},
  booktitle={Proceedings of the IEEE/CVF Conference on Computer Vision and Pattern Recognition},
  pages={26181--26190},
  year={2024}
}

@inproceedings{li2021mfes,
  title={Mfes-hb: Efficient hyperband with multi-fidelity quality measurements},
  author={Li, Yang and Shen, Yu and Jiang, Jiawei and Gao, Jinyang and Zhang, Ce and Cui, Bin},
  booktitle={Proceedings of the AAAI Conference on Artificial Intelligence},
  volume={35},
  number={10},
  pages={8491--8500},
  year={2021}
}

@inproceedings{jamieson2016non,
  title={Non-stochastic best arm identification and hyperparameter optimization},
  author={Jamieson, Kevin and Talwalkar, Ameet},
  booktitle={Artificial intelligence and statistics},
  pages={240--248},
  year={2016},
  organization={PMLR}
}

@inproceedings{hu2019multi,
  title={Multi-fidelity automatic hyper-parameter tuning via transfer series expansion},
  author={Hu, Yi-Qi and Yu, Yang and Tu, Wei-Wei and Yang, Qiang and Chen, Yuqiang and Dai, Wenyuan},
  booktitle={Proceedings of the AAAI Conference on Artificial Intelligence},
  volume={33},
  number={01},
  pages={3846--3853},
  year={2019}
}

@inproceedings{klein2017fast,
  title={Fast bayesian optimization of machine learning hyperparameters on large datasets},
  author={Klein, Aaron and Falkner, Stefan and Bartels, Simon and Hennig, Philipp and Hutter, Frank},
  booktitle={Artificial intelligence and statistics},
  pages={528--536},
  year={2017},
  organization={PMLR}
}

@article{li2018massively,
  title={Massively parallel hyperparameter tuning},
  author={Li, Lisha and Jamieson, Kevin and Rostamizadeh, Afshin and Gonina, Katya and Hardt, Moritz and Recht, Benjamin and Talwalkar, Ameet},
  year={2018}
}

@inproceedings{salinas2023optimizing,
  title={Optimizing hyperparameters with conformal quantile regression},
  author={Salinas, David and Golebiowski, Jacek and Klein, Aaron and Seeger, Matthias and Archambeau, Cedric},
  booktitle={International Conference on Machine Learning},
  pages={29876--29893},
  year={2023},
  organization={PMLR}
}

@article{klein2020model,
  title={Model-based asynchronous hyperparameter and neural architecture search},
  author={Klein, Aaron and Tiao, Louis C and Lienart, Thibaut and Archambeau, Cedric and Seeger, Matthias},
  journal={arXiv preprint arXiv:2003.10865},
  year={2020}
}

@article{mohr2024learning,
  title={Learning curves for decision making in supervised machine learning: a survey},
  author={Mohr, Felix and van Rijn, Jan N},
  journal={Machine Learning},
  volume={113},
  number={11},
  pages={8371--8425},
  year={2024},
  publisher={Springer}
}

@inproceedings{wistuba2020learning,
  title={Learning to rank learning curves},
  author={Wistuba, Martin and Pedapati, Tejaswini},
  booktitle={International Conference on Machine Learning},
  pages={10303--10312},
  year={2020},
  organization={PMLR}
}

@inproceedings{klein2017learning,
  title={Learning curve prediction with Bayesian neural networks},
  author={Klein, Aaron and Falkner, Stefan and Springenberg, Jost Tobias and Hutter, Frank},
  booktitle={International conference on learning representations},
  year={2017}
}

@book{kimball2013data,
  title={The data warehouse toolkit: The definitive guide to dimensional modeling},
  author={Kimball, Ralph and Ross, Margy},
  year={2013},
  publisher={John Wiley \& Sons}
}

@book{huyen2022designing,
  title={Designing machine learning systems},
  author={Huyen, Chip},
  year={2022},
  publisher={" O'Reilly Media, Inc."}
}

@book{warren2015big,
  title={Big Data: Principles and best practices of scalable realtime data systems},
  author={Warren, James and Marz, Nathan},
  year={2015},
  publisher={Simon and Schuster}
}

@article{zhao2023automatic,
  title={Automatic database knob tuning: A survey},
  author={Zhao, Xinyang and Zhou, Xuanhe and Li, Guoliang},
  journal={IEEE Transactions on Knowledge and Data Engineering},
  volume={35},
  number={12},
  pages={12470--12490},
  year={2023},
  publisher={IEEE}
}

@article{huang2023survey,
  title={Survey on performance optimization for database systems},
  author={Huang, Shiyue and Qin, Yanzhao and Zhang, Xinyi and Tu, Yaofeng and Li, Zhongliang and Cui, Bin},
  journal={Science China Information Sciences},
  volume={66},
  number={2},
  pages={121102},
  year={2023},
  publisher={Springer}
}

@inproceedings{van2017automatic,
  title={Automatic database management system tuning through large-scale machine learning},
  author={Van Aken, Dana and Pavlo, Andrew and Gordon, Geoffrey J and Zhang, Bohan},
  booktitle={Proceedings of the 2017 ACM international conference on management of data},
  pages={1009--1024},
  year={2017}
}

@inproceedings{zhang2021restune,
  title={Restune: Resource oriented tuning boosted by meta-learning for cloud databases},
  author={Zhang, Xinyi and Wu, Hong and Chang, Zhuo and Jin, Shuowei and Tan, Jian and Li, Feifei and Zhang, Tieying and Cui, Bin},
  booktitle={Proceedings of the 2021 international conference on management of data},
  pages={2102--2114},
  year={2021}
}

@article{kanellis2022llamatune,
  title={LlamaTune: sample-efficient DBMS configuration tuning},
  author={Kanellis, Konstantinos and Ding, Cong and Kroth, Brian and M{\"u}ller, Andreas and Curino, Carlo and Venkataraman, Shivaram},
  journal={Proceedings of the VLDB Endowment},
  volume={15},
  number={11},
  pages={2953--2965},
  year={2022},
  publisher={VLDB Endowment}
}

@inproceedings{shen2025tune,
  title={A-Tune-Online: Efficient and QoS-Aware Online Configuration Tuning for Dynamic Workloads},
  author={Shen, Yu and Xu, Beicheng and Lu, Yupeng and Chen, Donghui and Jiang, Huaijun and Xie, Zhipeng and Fu, Senbo and Zhang, Nan and Ren, Yuxin and Jia, Ning and others},
  booktitle={2025 IEEE 41st International Conference on Data Engineering (ICDE)},
  pages={2161--2173},
  year={2025},
  organization={IEEE}
}

@inproceedings{zhang2022towards,
  title={Towards dynamic and safe configuration tuning for cloud databases},
  author={Zhang, Xinyi and Wu, Hong and Li, Yang and Tan, Jian and Li, Feifei and Cui, Bin},
  booktitle={Proceedings of the 2022 International Conference on Management of Data},
  pages={631--645},
  year={2022}
}

@article{zhang2023efficient,
  title={An efficient transfer learning based configuration adviser for database tuning},
  author={Zhang, Xinyi and Wu, Hong and Li, Yang and Tang, Zhengju and Tan, Jian and Li, Feifei and Cui, Bin},
  journal={Proceedings of the VLDB Endowment},
  volume={17},
  number={3},
  pages={539--552},
  year={2023},
  publisher={VLDB Endowment}
}

@inproceedings{wei2025toptune,
  title={TopTune: Tailored Optimization for Categorical and Continuous Knobs Towards Accelerated and Improved Database Performance Tuning},
  author={Wei, Rukai and Liu, Yu and Hou, Yufeng and Cui, Heng and Zhang, Yongqiang and Zhou, Ke},
  booktitle={2025 IEEE 41st International Conference on Data Engineering (ICDE)},
  pages={613--626},
  year={2025},
  organization={IEEE}
}

@inproceedings{zhang2019end,
  title={An end-to-end automatic cloud database tuning system using deep reinforcement learning},
  author={Zhang, Ji and Liu, Yu and Zhou, Ke and Li, Guoliang and Xiao, Zhili and Cheng, Bin and Xing, Jiashu and Wang, Yangtao and Cheng, Tianheng and Liu, Li and others},
  booktitle={Proceedings of the 2019 international conference on management of data},
  pages={415--432},
  year={2019}
}

@article{li2019qtune,
  title={Qtune: A query-aware database tuning system with deep reinforcement learning},
  author={Li, Guoliang and Zhou, Xuanhe and Li, Shifu and Gao, Bo},
  journal={Proceedings of the VLDB Endowment},
  volume={12},
  number={12},
  pages={2118--2130},
  year={2019},
  publisher={VLDB Endowment}
}

@inproceedings{cai2022hunter,
  title={HUNTER: an online cloud database hybrid tuning system for personalized requirements},
  author={Cai, Baoqing and Liu, Yu and Zhang, Ce and Zhang, Guangyu and Zhou, Ke and Liu, Li and Li, Chunhua and Cheng, Bin and Yang, Jie and Xing, Jiashu},
  booktitle={Proceedings of the 2022 International Conference on Management of Data},
  pages={646--659},
  year={2022}
}

@inproceedings{wang2022udo,
  title={UDO: Universal Database Optimization using Reinforcement Learning},
  author={Wang, Junxiong and Trummer, Immanuel and Basu, Debabrota},
  booktitle={Proceedings of the VLDB Endowment},
  volume={14},
  number={13},
  pages={3402--3414},
  year={2022}
}

@inproceedings{sun2025rabbit,
  title={Rabbit: Retrieval-Augmented Generation Enables Better Automatic Database Knob Tuning},
  author={Sun, Wenwen and Pan, Zhicheng and Hu, Zirui and Liu, Yu and Yang, Chengcheng and Zhang, Rong and Zhou, Xuan},
  booktitle={2025 IEEE 41st International Conference on Data Engineering (ICDE)},
  pages={3807--3820},
  year={2025},
  organization={IEEE}
}

@article{lao2025gptuner,
  title={GPTuner: An LLM-Based Database Tuning System},
  author={Lao, Jiale and Wang, Yibo and Li, Yufei and Wang, Jianping and Zhang, Yunjia and Cheng, Zhiyuan and Chen, Wanghu and Tang, Mingjie and Wang, Jianguo},
  journal={ACM SIGMOD Record},
  volume={54},
  number={1},
  pages={101--110},
  year={2025},
  publisher={ACM New York, NY, USA}
}

@inproceedings{trummer2022db,
  title={DB-BERT: a Database Tuning Tool that" Reads the Manual"},
  author={Trummer, Immanuel},
  booktitle={Proceedings of the 2022 international conference on management of data},
  pages={190--203},
  year={2022}
}

@article{li2025agenttune,
  title={AgentTune: An Agent-Based Large Language Model Framework for Database Knob Tuning},
  author={Li, Yiyan and Li, Haoyang and Zhang, Jing and Borovica-Gajic, Renata and Wang, Shuai and Zhang, Tieying and Chen, Jianjun and Shi, Rui and Li, Cuiping and Chen, Hong},
  journal={Proceedings of the ACM on Management of Data},
  volume={3},
  number={6},
  pages={1--29},
  year={2025},
  publisher={ACM New York, NY, USA}
}

@inproceedings{alipourfard2017cherrypick,
  title={$\{$CherryPick$\}$: Adaptively unearthing the best cloud configurations for big data analytics},
  author={Alipourfard, Omid and Liu, Hongqiang Harry and Chen, Jianshu and Venkataraman, Shivaram and Yu, Minlan and Zhang, Ming},
  booktitle={14th USENIX Symposium on Networked Systems Design and Implementation (NSDI 17)},
  pages={469--482},
  year={2017}
}

@inproceedings{xin2022locat,
  title={Locat: Low-overhead online configuration auto-tuning of spark sql applications},
  author={Xin, Jinhan and Hwang, Kai and Yu, Zhibin},
  booktitle={Proceedings of the 2022 International Conference on Management of Data},
  pages={674--684},
  year={2022}
}

@article{chen2024tie,
  title={TIE: Fast experiment-driven ml-based configuration tuning for in-memory data analytics},
  author={Chen, Chao and Xin, Jinhan and Yu, Zhibin},
  journal={IEEE Transactions on Computers},
  volume={73},
  number={5},
  pages={1233--1247},
  year={2024},
  publisher={IEEE}
}

@inproceedings{fekry2020tune,
  title={To tune or not to tune? in search of optimal configurations for data analytics},
  author={Fekry, Ayat and Carata, Lucian and Pasquier, Thomas and Rice, Andrew and Hopper, Andy},
  booktitle={Proceedings of the 26th ACM SIGKDD International Conference on Knowledge Discovery \& Data Mining},
  pages={2494--2504},
  year={2020}
}

@article{li2023towards,
  title={Towards General and Efficient Online Tuning for Spark},
  author={Li, Yang and Jiang, Huaijun and Shen, Yu and Fang, Yide and Yang, Xiaofeng and Huang, Danqing and Zhang, Xinyi and Zhang, Wentao and Zhang, Ce and Chen, Peng and others},
  journal={Proceedings of the VLDB Endowment},
  volume={16},
  number={12},
  pages={3570--3583},
  year={2023},
  publisher={VLDB Endowment}
}

@inproceedings{shen2023rover,
  title={Rover: An online Spark SQL tuning service via generalized transfer learning},
  author={Shen, Yu and Ren, Xinyuyang and Lu, Yupeng and Jiang, Huaijun and Xu, Huanyong and Peng, Di and Li, Yang and Zhang, Wentao and Cui, Bin},
  booktitle={Proceedings of the 29th ACM SIGKDD Conference on Knowledge Discovery and Data Mining},
  pages={4800--4812},
  year={2023}
}

@article{li2025loftune,
  title={LOFTune: A Low-Overhead and Flexible Approach for Spark SQL Configuration Tuning},
  author={Li, Jiahui and Ye, Junhao and Mao, Yuren and Gao, Yunjun and Chen, Lu},
  journal={IEEE Transactions on Knowledge and Data Engineering},
  year={2025},
  publisher={IEEE}
}

@article{tibshirani1996regression,
  title={Regression shrinkage and selection via the lasso},
  author={Tibshirani, Robert},
  journal={Journal of the Royal Statistical Society Series B: Statistical Methodology},
  volume={58},
  number={1},
  pages={267--288},
  year={1996},
  publisher={Oxford University Press}
}

@article{lundberg2017unified,
  title={A unified approach to interpreting model predictions},
  author={Lundberg, Scott M and Lee, Su-In},
  journal={Advances in neural information processing systems},
  volume={30},
  year={2017}
}

@article{letham2020re,
  title={Re-examining linear embeddings for high-dimensional Bayesian optimization},
  author={Letham, Ben and Calandra, Roberto and Rai, Akshara and Bakshy, Eytan},
  journal={Advances in neural information processing systems},
  volume={33},
  pages={1546--1558},
  year={2020}
}

@inproceedings{nayebi2019framework,
  title={A framework for Bayesian optimization in embedded subspaces},
  author={Nayebi, Amin and Munteanu, Alexander and Poloczek, Matthias},
  booktitle={International Conference on Machine Learning},
  pages={4752--4761},
  year={2019},
  organization={PMLR}
}

@article{parzen1962estimation,
  title={On estimation of a probability density function and mode},
  author={Parzen, Emanuel},
  journal={The annals of mathematical statistics},
  volume={33},
  number={3},
  pages={1065--1076},
  year={1962},
  publisher={JSTOR}
}

@online{openEuler,
    author =        "OpenEuler",
    url =          "https://www.openeuler.org/",
    title =        "An Open Source OS for Digital Infrastructure",
    year =         "2025",
    note =         "Last accessed: July 5, 2026"
}

@article{jiang2024openbox,
  title={Openbox: A Python toolkit for generalized black-box optimization},
  author={Jiang, Huaijun and Shen, Yu and Li, Yang and Xu, Beicheng and Du, Sixian and Zhang, Wentao and Zhang, Ce and Cui, Bin},
  journal={Journal of Machine Learning Research},
  volume={25},
  number={120},
  pages={1--11},
  year={2024}
}

@article{hive_thusoo2009hive,
  title={Hive: a warehousing solution over a map-reduce framework},
  author={Thusoo, Ashish and Sarma, Joydeep Sen and Jain, Namit and Shao, Zheng and Chakka, Prasad and Anthony, Suresh and Liu, Hao and Wyckoff, Pete and Murthy, Raghotham},
  journal={Proceedings of the VLDB Endowment},
  volume={2},
  number={2},
  pages={1626--1629},
  year={2009},
  publisher={VLDB Endowment}
}

@inproceedings{presto_sethi2019presto,
  title={Presto: SQL on everything},
  author={Sethi, Raghav and Traverso, Martin and Sundstrom, Dain and Phillips, David and Xie, Wenlei and Sun, Yutian and Yegitbasi, Nezih and Jin, Haozhun and Hwang, Eric and Shingte, Nileema and others},
  booktitle={2019 IEEE 35th International Conference on Data Engineering (ICDE)},
  pages={1802--1813},
  year={2019},
  organization={IEEE}
}

@inproceedings{spark_zaharia2010spark,
  title={Spark: Cluster computing with working sets},
  author={Zaharia, Matei and Chowdhury, Mosharaf and Franklin, Michael J and Shenker, Scott and Stoica, Ion},
  booktitle={2nd USENIX workshop on hot topics in cloud computing (HotCloud 10)},
  year={2010}
}

@article{box_perrone2019learning,
  title={Learning search spaces for bayesian optimization: Another view of hyperparameter transfer learning},
  author={Perrone, Valerio and Shen, Huibin and Seeger, Matthias W and Archambeau, Cedric and Jenatton, Rodolphe},
  journal={Advances in neural information processing systems},
  volume={32},
  year={2019}
}

@inproceedings{siddiqui2022isum,
  title={ISUM: Efficiently compressing large and complex workloads for scalable index tuning},
  author={Siddiqui, Tarique and Jo, Saehan and Wu, Wentao and Wang, Chi and Narasayya, Vivek and Chaudhuri, Surajit},
  booktitle={Proceedings of the 2022 International Conference on Management of Data},
  pages={660--673},
  year={2022}
}

@article{deep2020comprehensive,
  title={Comprehensive and efficient workload compression},
  author={Deep, Shaleen and Gruenheid, Anja and Koutris, Paraschos and Naughton, Jeffrey and Viglas, Stratis},
  journal={Proceedings of the VLDB Endowment},
  volume={14},
  number={3},
  pages={418--430},
  year={2020},
  publisher={VLDB Endowment}
}

@inproceedings{chaudhuri2002compressing,
  title={Compressing sql workloads},
  author={Chaudhuri, Surajit and Gupta, Ashish Kumar and Narasayya, Vivek},
  booktitle={Proceedings of the 2002 ACM SIGMOD international conference on Management of data},
  pages={488--499},
  year={2002}
}

@inproceedings{chang2024mfix,
  title={Mfix: An efficient and reliable index advisor via multi-fidelity bayesian optimization},
  author={Chang, Zhuo and Zhang, Xinyi and Li, Yang and Miao, Xupeng and Qin, Yanzhao and Cui, Bin},
  booktitle={2024 IEEE 40th international conference on data engineering (ICDE)},
  pages={4343--4356},
  year={2024},
  organization={IEEE}
}

@inproceedings{zhu2025rockhopper,
  title={Rockhopper: A robust optimizer for spark configuration tuning in production environment},
  author={Zhu, Yiwen and Sen, Rathijit and Kroth, Brian and Matusevych, Sergiy and Mueller, Andreas and Huang, Tengfei and Challapalli, Rahul and Tang, Weihan and He, Xin and Liu, Mo and others},
  booktitle={Companion of the 2025 International Conference on Management of Data},
  pages={743--756},
  year={2025}
}

@inproceedings{freischuetz2025tuna,
  title={Tuna: Tuning unstable and noisy cloud applications},
  author={Freischuetz, Johannes and Kanellis, Konstantinos and Kroth, Brian and Venkataraman, Shivaram},
  booktitle={Proceedings of the Twentieth European Conference on Computer Systems},
  pages={954--973},
  year={2025}
}

@article{ding2021dsb,
  title={DSB: A decision support benchmark for workload-driven and traditional database systems},
  author={Ding, Bailu and Chaudhuri, Surajit and Gehrke, Johannes and Narasayya, Vivek R.},
  journal={Proceedings of the VLDB Endowment},
  volume={14},
  number={13},
  pages={3376--3388},
  year={2021},
  publisher={VLDB Endowment}
}

@article{zhu2023lero,
  title={Lero: A learning-to-rank query optimizer},
  author={Zhu, Rong and Chen, Wei and Ding, Bolin and Chen, Xingguang and Pfadler, Andreas and Wu, Ziniu and Zhou, Jingren},
  journal={arXiv preprint arXiv:2302.06873},
  year={2023}
}

@article{jin2025postman,
  title={PostMan: A Productive System for Spatio-temporal Data Management and Analysis: J. Jin et al.},
  author={Jin, Jiaqi and Fang, Ziquan and Chen, Lu and Gao, Yunjun},
  journal={Data Science and Engineering},
  pages={1--24},
  year={2025},
  publisher={Springer}
}

@article{frazzetto2025graph,
  title={Graph Neural Networks for Candidate-Job Matching: An Inductive Learning Approach: P. Frazzetto et al.},
  author={Frazzetto, Paolo and Haq, Muhammad Uzair Ul and Fabris, Flavia and Sperduti, Alessandro},
  journal={Data Science and Engineering},
  pages={1--18},
  year={2025},
  publisher={Springer}
}

@article{li2024opengauss,
  title={openGauss: An enterprise-grade open-source database system},
  author={Li, Guo-Liang and Wang, Jiang and Chen, Guo},
  journal={Journal of Computer Science and Technology},
  volume={39},
  number={5},
  pages={1007--1028},
  year={2024},
  publisher={Springer}
}

@article{li2024sha,
  title={Sha: Qos-aware software and hardware auto-tuning for database systems},
  author={Li, Jin and Chen, Quan and Tang, Xiao-Xin and Guo, Min-Yi},
  journal={Journal of Computer Science and Technology},
  volume={39},
  number={2},
  pages={369--383},
  year={2024},
  publisher={Springer}
}
